\documentclass[epj]{svjour}
\usepackage{graphics}
\usepackage[utf8]{inputenc}
\usepackage{multicol}
\usepackage{tabularx} 
\usepackage{lscape}
\usepackage{booktabs} 
\usepackage{hyperref}
\usepackage{grffile} 
\usepackage{array}
\usepackage{framed}
\usepackage{color}
\usepackage{bm,enumerate,amsmath,amssymb}
\usepackage{etoolbox}
\usepackage{comment}

\usepackage{epigraph}
\usepackage{amsmath}
\usepackage{physics}
\usepackage{mathtools, cuted}
\usepackage{siunitx}
\usepackage{stackengine} 
\usepackage{epsfig}
\usepackage{float} 
\usepackage[colorinlistoftodos, textwidth=20mm]{todonotes}
\newcommand{\eq}[1]{\begin{equation}{#1}\end{equation}}

\begin{document}
\title{Challenges in modelling diffusiophoretic transport}
\author{Simón Ramírez-Hinestrosa \and Daan Frenkel}
\institute{Yusuf Hamied Department of Chemistry, University of Cambridge, UK}
\date{Received: date / Revised version: date}
%
\abstract{The methodology to simulate transport phenomena in bulk systems is well-established.   
In contrast, there is no clear consensus about the choice of techniques to model cross-transport phenomena and phoretic transport, mainly because some of the hydrodynamic descriptions are incomplete from a thermodynamic point of view. 
In the present paper, we use a unified framework to describe diffusio-osmosis(phoresis), and we report non-equilibrium Molecular Dynamics (NEMD) on such systems.
We explore different simulation methods to highlight some of the technical problems that arise in the calculations.
For diffusiophoresis, we use two NEMD methods: boundary-driven and field-driven. Although the two methods should be equivalent in the limit of very weak gradients, we find that finite Peclet-number effects are much stronger in boundary-driven flows than in the case where we apply fictitious color forces. 
}
%
\maketitle
\section{Introduction}
Chemical potential gradients in a bulk fluid cannot cause flow, as they do not result in net forces on any sub-volume of the fluid. 
The reason is that there are two ways in which the momentum of a fluid element can change: 
1) due to a net externally applied force (e.g. gravity) on the particles in the volume and 
2) due to a net imbalance between momentum flowing in through opposing boundaries of the volume element.
But momentum flux through a boundary is what we normally call pressure. Therefore, an imbalance of momentum flux through opposing boundaries results if the pressure were not uniform.

If we consider a bulk fluid at constant pressure, and in the absence of external forces, other thermodynamic driving forces, such as gradients in $T$ or $\mu$, cannot cause net forces on a fluid element.

To illustrate this, consider a bulk binary system composed of $N_f$ solvent particles $f$ and $N_s$ solute particles $s$. 

We assume that the composition is not homogeneous. 
Then each species is subject to a chemical potential gradient $\grad \mu_i$, for $i=s,f$. 
We consider the case that the pressure in the bulk of the fluid is constant, and for simplicity, we also assume that the temperature is constant.
Although the system as a whole is not in equilibrium, we assume local thermodynamic equilibrium. 
We can then write the Gibbs-Duhem relation as
\eq{\label{GibbsDuhem} V\;dP-S\;dT=N_s\;d\mu_s\ + N_s\;d\mu_f = 0;}
which, at constant $P$ and $T$, implies:  
\eq{\label{Force_in_bulk}N_f\grad \mu_f = - N_f \grad \mu_s\;.}

It is often convenient to interpret a gradient in the chemical potential of species $i$ as (minus) a force that acts on this species.
The introduction of such fictitious, species-dependent ``color'' forces is allowed because the gradient of a chemical potential has the same effect as the gradient of a real potential acting on a given species.
This is, of course, well-known for electrolyte solutions where gradients of the electrostatic potential and the chemical potential have the same effect.

Importantly, the Gibbs-Duhem equation (\ref{GibbsDuhem}) establishes a relation between the color forces:
if each particle of species $i$ is subject to a color force $F_i \sim -\grad \mu_i$, then Eq.~\eqref{Force_in_bulk} expresses the fact that the net force on a fluid element vanishes.

\begin{figure}[H]
\centering
\includegraphics[width=0.6\linewidth]{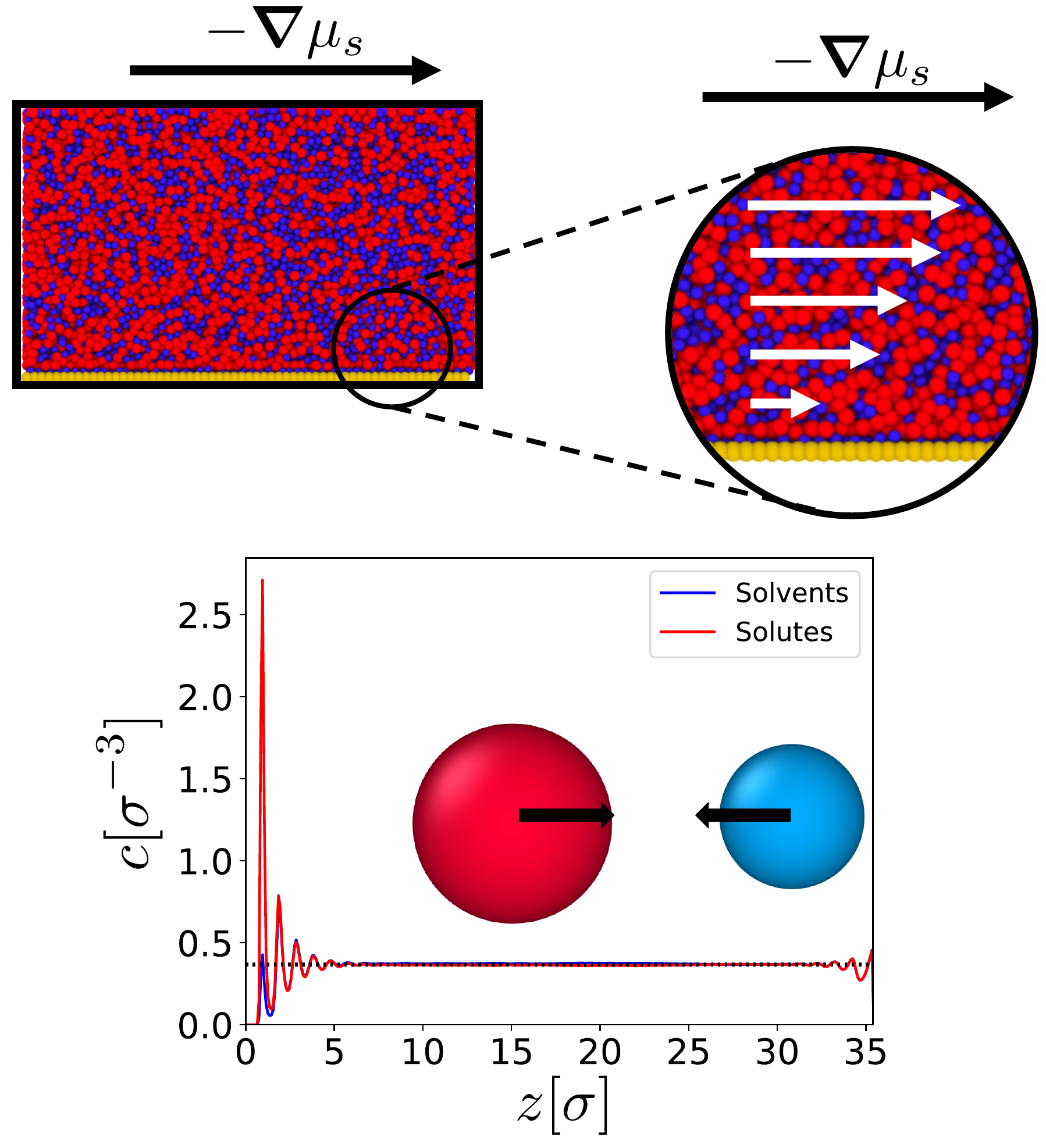}
\caption{The preferential interaction of the solutes with a solid surface creates an excess of this species at the interface. The thermodynamic force $-\grad \mu_s$ drives the solute motion creating a net flux due to the excess at the interface, defining the flow of the whole system.}  
\label{fig:microscopic_origin}
\end{figure}

However, contrary to what happens in the bulk, a gradient in the chemical potential of the various components in a fluid mixture can cause a net hydrodynamic flow in the presence of an interface that interacts differently with the different species in the solution. In Fig.~\ref{fig:microscopic_origin}, we show a flat solid wall and a binary solution composed of solutes $s$ and solvents $f$.
Each species interacts with the wall differently, with solutes being adsorbed preferentially at the solid surface. 
The adsorption creates an excess of solutes in the diffuse layer.
Moreover, if there is a chemical potential gradient on the solutes $\grad \mu_s$, then they move following the thermodynamic force $-\grad \mu_s$. 
As a result of the excess at the interface, the solute movement drives the solution flow. 
All this takes place within the diffuse layer, beyond which the fluid moves force-free; thus we observe the typical plug-like flow \cite{Yoshida2017,Liu2018}.
Such a flow, induced by chemical-potential gradients, is known as diffusio-osmosis. 
Other flows that are enabled by the presence of an interface are electro-osmosis and thermo-osmosis, each one having an ``excess" quantity associated. 
The former originating from an excess of charges and the latter from the excess enthalpy at the interface. 

Surface-induced, ``phoretic'' flow phenomena are usually negligible  in macroscopic channels, but can become dominant in micro or nano-scale channels, as phoretic fluxes scale as the channel diameter squared, whereas Poisseuille fluxes scale as the fourth power.
From now on, we will often use the term ``phoretic'' transport to the wider class of surface-induced flow phenomena, even though, strictly speaking, phoresis is the phenomenon where particles move under the influence of the same gradients that can cause flow along fixed surfaces. 

Simulations provide a tool to gain a better microscopic understanding of the factors that affect phoretic flows. 
In particular, simulations could make it possible to {\em predict} the strength of such flows based on the knowledge of the relevant intermolecular interactions.
This in contrast to the more traditional descriptions that make use of hydrodynamic continuum theory and thermodynamics. Clearly, the need for quantitative understanding of phoretic transport is growing as more research focuses on nano-scale phenomena.  
But simulations of phoretic transport require special care, as they require approaches that differ from their bulk counterparts.
Over the past years, much progress in this direction has been made.
In this paper, we focus on one particular form of phoretic transport, namely diffusio-osmotic flow.

\begin{figure}[H]
\centering
\includegraphics[width=0.8\linewidth]{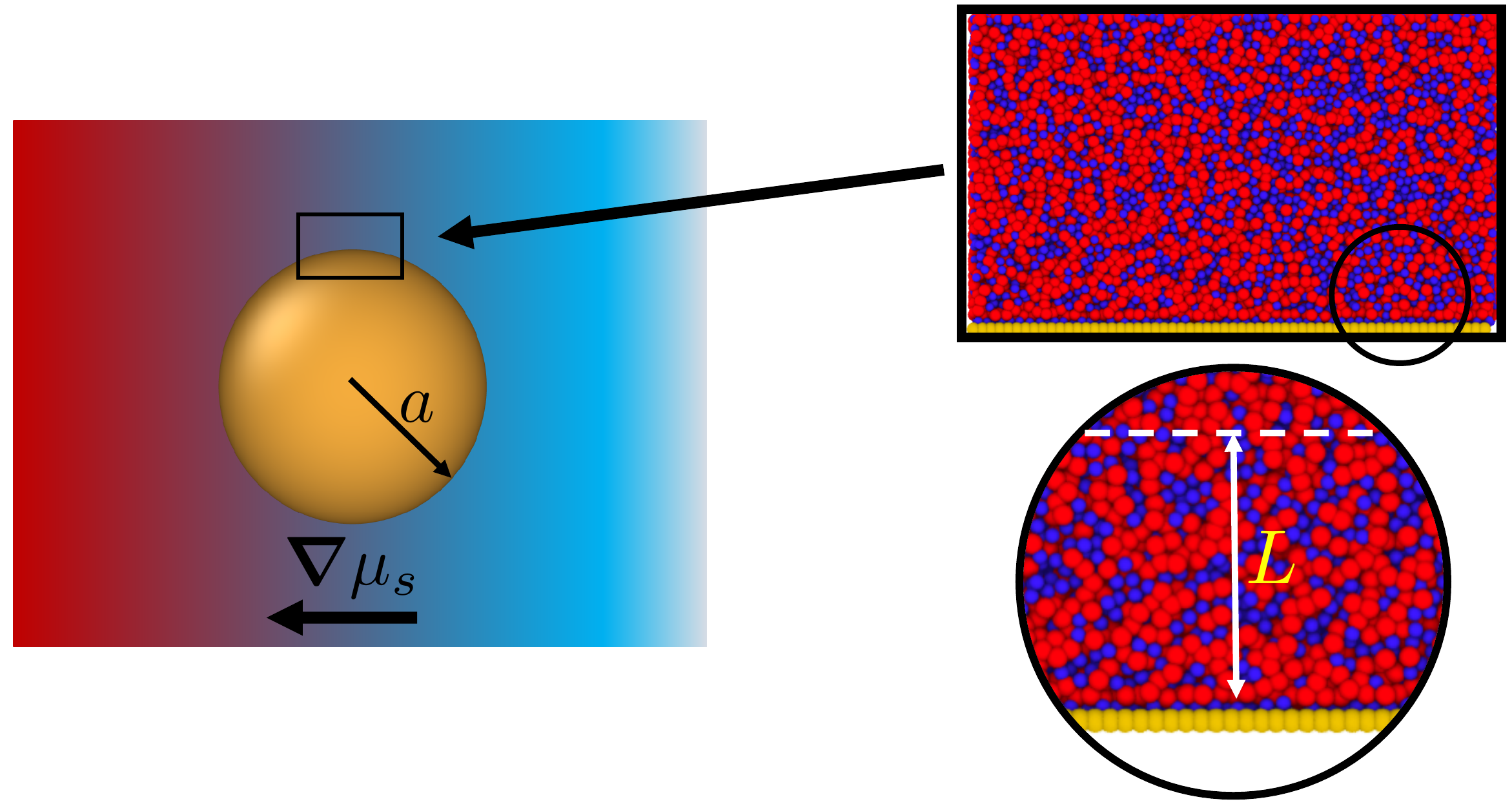}
\caption{Diffusio-osmosis can be seen as diffusiophoresis under the boundary layer approximation. 
Rather than focusing on the movement of the colloidal particle, we focus on the fluid flow on its surface. 
In the case that $a\gg L$,  this reduces the analysis to a fluid flow on top of a flat plate, known as Derjaguin's approximation. }  
\label{fig:Multiscale}
\end{figure}

Diffusio-osmotic flow is a subject that was introduced by Derjaguin, using the language of thermodynamics and hydrodynamics. 
As an example, the presence of the colloid perturbs the neighbouring fluid creating a heterogeneous region close to its surface known as the diffuse layer. 
We consider the case that the colloid radius $a$ is much larger than the thickness $L$ of the diffuse layer. 
Derjaguin introduced this ``boundary layer approximation'' \cite{Derjaguin1947}, to separate the problem into two regions: one inside and the other  outside the diffuse layer. 
Due to the scale separation, the dynamics can be studied inside the diffusive layer. 
In this approximation, the diffusio-phoretic problem reduces to studying  the flow of a fluid induced by a gradient of chemical potential parallel to  a flat surface (see  Fig.~\ref{fig:Multiscale}. 

As was already noted by Barrat and Bocquet~\cite{Barrat1999}, a continuum approach is perfectly adequate to describe the hydrodynamics of fluids at a distance of more than a few molecular diameters from a surface.
However, in order to estimate the magnitude of the velocity profile close to the surface, a microscopic picture is needed.
In a sense, this fact is already clear from the long discussion about the meaning of the $\zeta$-potential in electro-kinetic flows: this quantity depends sensitively on the local fluidity and molecular arrangements near a surface and typically cannot be predicted with any accuracy on the basis of macroscopic arguments alone.
Moreover, even though the action of the surface is usually very local, its effect extends into the bulk as it changes the effective hydrodynamic boundary conditions~\cite{Bocquet2009}.

Early Molecular Dynamics (MD) simulations to investigate diffusio-osmosis were performed by Ajdari \& Bocquet \cite{Ajdari2006}, who made use of the Onsager relations (see e.g~\cite{DeGroot1984}) to measure the diffusio-osmotic transport coefficient indirectly by measuring the excess solute flux due to an applied  pressure gradient. 
More recent simulations have used both equilibrium and non-equilibrium MD  techniques to study diffusio-osmosis~\cite{Yoshida2017,Liu2017c,Liu2018,Mangaud2020}. 
In addition, there have been several reports on MD simulations of diffusiophoresis of colloids~\cite{Sharifi-Mood2013a,Wei2020} and short polymers~\cite{Ramirez-Hinestrosa2019}. 

In this article, we aim to discuss the existing theories and MD methods to study diffusio-osmosis(phoresis) in a unified framework.
 Our first case study is the diffusio-osmotic flow in simple planar geometry. 
 We first derive the expression for the entropy production for the transport driven by chemical potential gradients, using non-equilibrium thermodynamics. 
 A crucial step is to construct a consistent set of thermodynamic forces and fluxes, which allows finding the Green-Kubo expressions for the diffusio-osmotic transport coefficient. We perform simulations using Non-Equilibrium molecular dynamics applying microscopic forces that represent the effect of chemical potential gradients. Moreover, we propose an alternative route to derive a theoretical estimate for the diffusio-osmotic flow velocity. Our general expression reduces to the well-known theoretical results by Derjaguin \cite{Anderson1989} and Anderson \cite{Derjaguin1947} in the limit of an ideal-dilute solution in the bulk.

Later, we study colloidal diffusiophoresis. We examine a spherical particle under the influence of solutes in a binary solution. We performed simulations using two non-equilibrium techniques. We first imposed the explicit thermodynamic force driving the phoretic motion. Alternatively, we used the microscopic representation of the chemical potential gradient. We show that the hydrodynamic regime given by the Peclet number is crucially different for the two approaches.

\section{Simulation  techniques}\label{sec:SimulationTechniques}
Phoretic transport occurs in systems out of equilibrium, but if the applied driving forces are small enough, it is possible to estimate transport coefficients using linear response theory~\cite{Yoshida2017,Mangaud2020}. 
The main advantage of working in the linear regime is that it allows us to compute transport coefficients by studying auto or cross time-correlation in equilibrium. 
As was shown by Onsager~\cite{ons311}, the transport matrix that provides the linear relation between the fluxes $J_i$ and the (thermodynamic) forces $X_j$ is symmetrical:

\[
J_i=\sum_j M_{ij} X_j \;,
\]
with $M_{ij}=M_{ji}$. 
\begin{figure}[H]
\centering
\includegraphics[width=\linewidth]{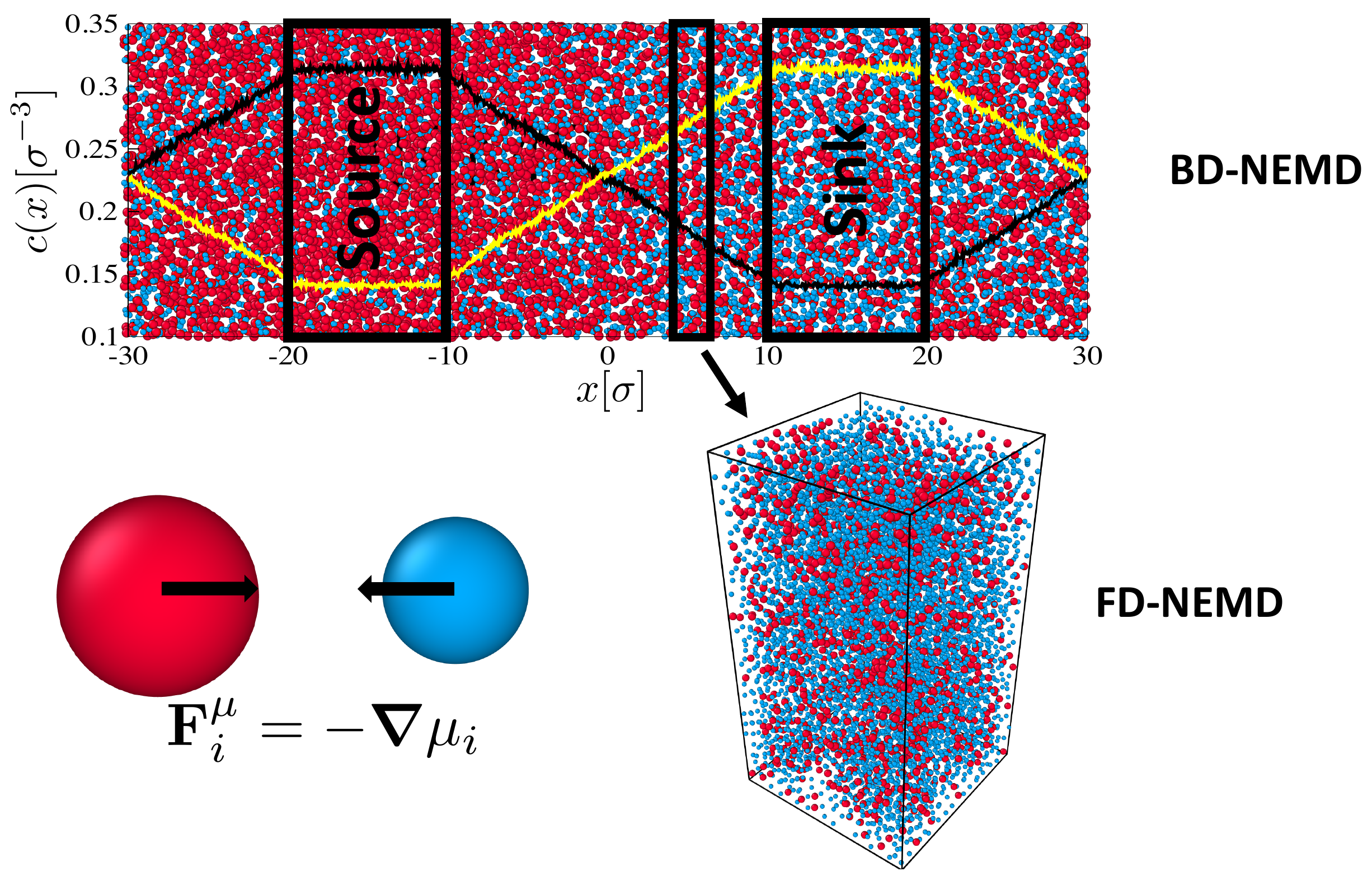}
\caption{Bulk diffusion in an ideal binary solution at constant temperature and pressure. We show a BD-NEMD simulation box with \textit{source} and \textit{sink} regions where we impose, respectively, a high and a low concentration of the red particles. The FD-NEMD box can be viewed as representing the same thermodynamic state as a small region in local thermodynamic equilibrium (LTE) from the BD-NEMD simulation. 
In FD-NEMD, the concentration is homogeneous and a force representing the effect the chemical potential gradient is applied to each particle. }  
\label{fig:BD_to_EF}
\end{figure} 
\subsection{Non-Equilibrium Molecular Dynamics (NEMD)} \label{sec:ch_bg_NEMD}
In real systems, phoretic transport is the result of some externally imposed gradient in the thermodynamic fields (temperature, chemical or electrical potential) that determine the equilibrium properties of a system.

In a simulation, one can choose to impose such inhomogeneities, but alternatively, one can apply a fictitious external field that has the same effect as these inhomogeneities. 
The idea behind this approach becomes clear if we consider Einstein's derivation of the relation between the diffusion coefficient $D$ and the mobility $m$ of a particle~\cite{ein051}: $ m=D/k_BT$.
Einstein considered the balance between the flux of particles under the influence of a concentration gradient and the counterbalancing flux due to an external potential gradient.

The advantage of using an external field, rather than the original concentration gradient, is that the field can be kept constant in a system with periodic boundary conditions, whereas gradients due to real variations in the concentration must be periodic, to be compatible with the boundary conditions.
Furthermore, we assume that different species can be driven by different fields. 
Fields that act specifically on particles of a given type only, are usually called {\em color forces}.
Although the results of the calculations should not depend on which approach is chosen, we shall see later that, in the non-linear regimes, interesting differences appear. 

In what follows, we shall refer to simulations where the fluxes are due to periodically repeated concentration differences, as Boundary-Driven Non-equilibrium MD (BD-NEMD). 
If the fluxes are due to constant color fields, we will refer to Field-Driven Non-Equilibrium MD (FD-NEMD). 

We note that the gradients could also be created between two large but finite reservoirs, in which case periodic boundary conditions would not be needed, provided the conditions in the reservoirs are being kept fixed.
However, without periodic boundary conditions, net flow along the direction of the gradients is not possible.

In BD-NEMD we create two spatially separated reservoirs in the simulation box:  typically, two slabs separated by half the box-length in the $x$-direction.
The  particle concentrations in these two slabs are fixed at different values, to maintain a concentration gradient.
In FD-NEMD we apply an external color force on each particle, which mimics the influence of the thermodynamic force. 
In Fig.~\ref{fig:BD_to_EF}, we show the connection between BD-NEMD and FD-NEMD simulations of bulk diffusion.
\subsection{Boundary-Driven Non-equilibrium Molecular Dynamics}
\label{sec:BD-NEMD}
The most intuitive way of imposing a chemical potential gradient in a simulation is to explicitly create two reservoirs in the simulations separated by a transport region as shown in Fig.~\ref{fig:BD_to_EF}. In this case, the concentrations at the boundary of the transport region, define the flux within, thus the name "boundary-driven".
The first simulations of systems experiencing chemical potential gradients in the context of diffusion were developed almost simultaneously by Heffelfinger and van Swol \cite{Heffelfinger1994} and MacElroy  \cite{MacElroy1994}. The former authors called the method Dual Control Volume Grand Canonical Molecular Dynamics (DCV-GCMD), as it consists of two grand canonical MC (GCMC) control volumes or reservoirs embedded in an MD-NVT simulation box. The GCMC serves to keep the desired concentration in the reservoirs. The molecules flow between the two control volumes, from the \textit{source} at high concentration to the \textit{sink} at a lower concentration. Replenishing the particles in the reservoirs at the right rate generates a steady-state flux of particles. 
This step is critical, as it may give incorrect results if the MC/MD frequency is not large enough \cite{Heffelfinger1994,Arya2001,Chempath2004}. The tuning depends on the size of the reservoirs, the distance between them and the number of GCMC insertion/deletion attempted per MC step.

BD-NEDM is inherently inhomogeneous. 
The approach is perfectly suited to simulate microscopically inhomogeneous systems, such as the flow through nanoscopic films  \cite{Thompson1999}. 
However, in other cases, the method has many disadvantages. 
As discussed before, it is difficult to tune the parameters to set up the initial concentration profile. 
Moreover, the use of GCMC implies that the velocity of the inserted particles must be known {\em a priori} and the method becomes problematic for fluid mixtures with large size ratio \cite{Thompson1999}. 
As we will discuss below, the magnitude of the gradient can lead to simulations occurring outside the linear response regime \cite{Arya2001}. 
Finally, the simulations tend to be time-consuming, as they must explicitly include the reservoirs, and there is an overhead associated with the MC insertion/deletions, or at the very least swaps of particle identities. 

\subsection{Field-Driven Non-Equilibrium Molecular Dynamics}
\label{sec:ch_bg_FD-NEMD}
Simulations using FD-NEMD require the introduction of an external field mimicking the effect of a thermodynamic force. In general, this \textit{synthetic} force has no clear physical interpretation, but its mechanical nature facilitates the simulation \cite{Evans1984} . 

The FD-NEMD approach has been applied extensively by~\cite{Evans2008} et al., with the body force coupling to particle variables such as the mass or the charge \cite{Todd2017}. 
In the case of diffusion, Maginn~{\em~et~al.}~\cite{Maginn1993} performed NEMD using a color field, in which particles are assigned color charges according to their chemical identity. In this way, they replaced the chemical potential gradient by a \textit{color force} of equal magnitude but opposite sign.

In practice, as the external field is non-conservative (i.e. as it is not the gradient of a potential), it will result in a constant dissipation in the system at steady state. 
Hence, the use of color fields must be combined with the use of a thermostat.
In what follows, we will make use of the  Nos\'e-Hoover thermostat, as it is a global thermostat, and it conserves linear momentum~\cite{Hoover1985}.

In 2001, Arya {\em et al.}~\cite{Arya2001} wrote about the use of color forces: ``this method has not been widely used, perhaps because the equivalence of such a homogeneous external forcing function that drives diffusion and an actual chemical potential gradient has not been formally demonstrated". 
However, subsequently, Yoshida {\em et al.}~\cite{Yoshida2017} justified replacing the imposed gradients with a constant color-force field on the basis of linear response theory, from which it also follows that the Onsager reciprocity relations hold for phoretic transport. 
Han {\em et al.}~\cite{Han2005} used a different method to simulate thermo-osmosis, 
which assumes that the forces on fluid elements can be computed from the gradient of the local, microscopic pressure tensor profile near a solid wall. 
However, as discussed in refs.~\cite{Liu2017c,Liu2018,Ganti2017}, the stress route is problematic in an inhomogeneous system (e.g. close to a wall) as the definition of the microscopic stress tensor is not unique.
Different definitions of the stress tensor lead to different estimates for the force and ultimately to different diffusio-osmotic flow velocities. 

To summarise, the advantages that FD-NEMD offers over BD-NEMD are that it allows the simulation the effect of a constant chemical potential gradient under periodic boundary conditions.
Moreover, we can use a homogeneous simulation box compatible with local thermal equilibrium. Lastly, we will show in Secs~\ref{sec:GK} and~\ref{section:colloid_FD} that the use of FD-NEMD makes it possible to explore also (mild) non-linear effects. 

\section{Diffusio-osmosis}
\label{Ch:Diffusio-osmosis}
Before starting the discussion, it is worth pointing out that in the literature, concentration and chemical potential gradients are taken as equivalent driving forces for diffusion. 
As concentration gradients are not proper thermodynamic driving forces, we will not use them, even though they are related to chemical potential gradients. To illustrate the difference: in an ideal solution, the driving force is proportional to the gradient in the logarithm of the concentration rather than a gradient in the concentration.
The language of chemical potential gradients is absolutely essential to take into account that not all gradients are independent, because of the Gibbs-Duhem relation.
In the language of concentration gradients, this effect is less obvious and often assumed to be negligible~\cite{Gupta2019} (for a discussion, see~\cite{Sanborn2001}). 

\subsection{Diffusio-osmosis and  entropy generation }
\begin{figure}[H]
\centering
\includegraphics[width=0.9\linewidth]{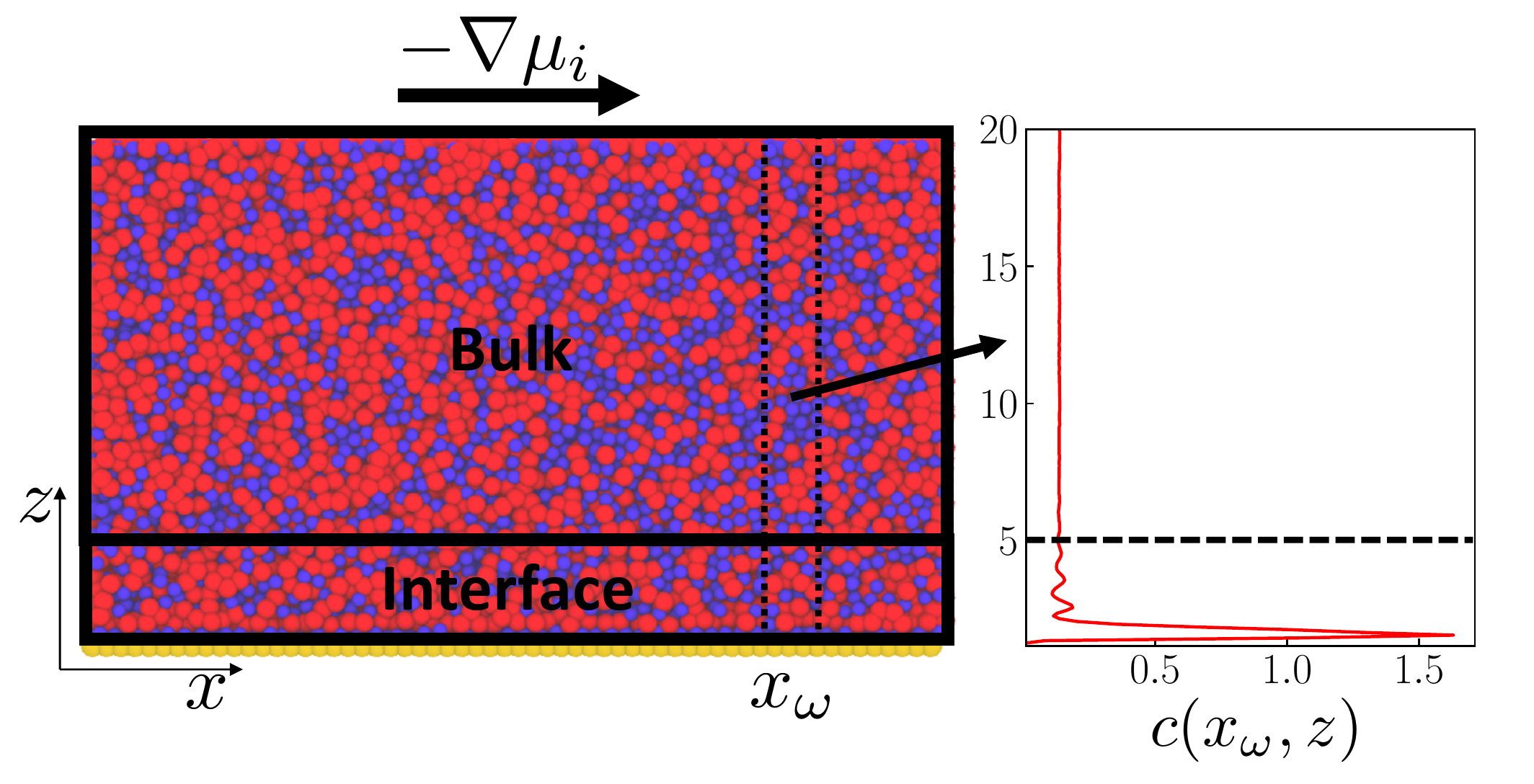}
\caption{The figure shows an $n$-component solution (only two are shown) in contact with a solid surface. 
The fluid is divided into two regions: bulk, where the concentration of the fluid $c(x,z)$ is independent of $z$, and the interface, where the local concentrations differ from their bulk values,  even in the absence of imposed concentration gradients. 
We consider the case that the gradient of the chemical potential ($\grad \mu_i $) is along the $x$-direction. A typical $z$-dependent density profile ($c(x_{\omega},z)$) of the fluid at a position $x_{\omega}$ is shown in the right-hand panel of the figure.}
\label{fig:DO_system}
\end{figure}

In our description of diffusio-osmosis, we consider a $n$-component fluid in contact with a solid surface, as shown in Fig.~\ref{fig:DO_system}. 
Initially, the only thermodynamic forces acting on the system are the chemical potential gradients of each species $i$, $\grad \mu_i$.
The fluid can be divided into two regions: the bulk, where the fluid can be considered homogeneous, and the vicinity of the (solid-liquid) interface, where the concentration of the different species at a distance $z$ from the interface, $c_i(z)$, differs from its bulk value. 
This deviation from the bulk concentrations decays as the distance from the surface is increased.
The reason why we first consider the expression for the entropy production is because it
contains both the thermodynamic driving forces and the conjugate fluxes~\cite{Prigogine1955}.

We start from the expression for the entropy production with no temperature gradient or chemical reactions~\cite{DeGroot1984},
\eq{\label{eq:dissipation_function}\Phi=T\sigma_s=\sum_{i=1}^N\mathbf{J_i}\cdot (-\grad \mu_i)\;.}
Here $\Phi$ is the dissipation function, which has units of energy density per unit of time. It is proportional to the rate of entropy production $\sigma_s$ and represents the dissipation of energy by an irreversible process in a control volume~\cite{Katzir-Katchalsky1965}.
The gradient in the chemical potential can be expressed as
\eq{\grad \mu_i\equiv(\grad \mu_i)_{P,T}+\left(\pdv{\mu_i}{P}\right)_{c_j,T}\nabla P\;,}
where $c_j$ indicates that the derivative is evaluated at constant concentration of the additional $n-1$ species with $j\ne i$. Additionally, we know that
\eq{\left(\pdv{\mu_i}{P}\right)_{c_j,T}=\nu_i\;,}
with $\nu_i$ being the partial molar volume of species $i$. 
Therefore, we can express the dissipation function in Eq~\eqref{eq:dissipation_function} as
\eq{\Phi=\left(\sum_{i=1}^n \mathbf{J_i} \nu_i \right) \cdot (-\grad P) + \sum_{i=1}^n \mathbf{J_i} \cdot (-\grad \mu_i)_{P,T}\;.}
The total volume flux in the system $\mathbf{Q}$ is defined as
\eq{\label{EPGeneraldt} \mathbf{Q} \equiv\sum_{i=1}^n \nu_i \mathbf{J}_i\;,} 
which is the average volume flow velocity in the system.
We can then express the dissipation function as
\eq{\label{PhiDependent}\Phi= \mathbf{Q} \cdot (-\grad P) + \sum_{i=1}^n \mathbf{J_i} \cdot (-\grad \mu_i)_{P,T}\;.}
The expression in Eq.~\eqref{PhiDependent} is convenient as it separates the diffusive fluxes, which  are Galilei-invariant, from the fluid flow, which is not. 
As we assume that in any infinitesimal volume element local equilibrium holds, we can use the Gibbs-Duhem relation,
\eq{\label{eq:Gibbs-Duhem}V dP =\sum_{i=1}^n N_i\;d\mu_i\;,}
where $N_i$ is the number of particles of species $n$.

Defining the densities $c_i\equiv N_i/V$, we can rewrite Eq.~\eqref{eq:Gibbs-Duhem} as:
\eq{\label{eq:GD_connection}\grad P =\sum_{i=1}^n c_i\;\grad\mu_i\;.}
Eq.~\eqref{eq:GD_connection} establishes a general relation between the thermodynamic forces in the system at constant temperature.
If we choose $(\grad \mu_i)_{P, T}$ in Eq.~\eqref{PhiDependent} as the independent driving forces then $\grad P$ is fixed. 
Conversely, if we use $\grad P$ as a driving force, then one of the $(\grad \mu_i)_{P, T}$ is linearly dependent on the others. The connection between thermodynamic forces (fluxes) avoids problems arising from treating them independently as discussed by Gupta~{\em et al.}~\cite{Gupta2019}. Note that the pressure that can be held constant in an experiment is the {\em bulk} pressure~\cite{DeGroot1984,Kjelstrup}.
If we impose a bulk pressure gradient, there will be fluid flow.
However, even when the pressure in the bulk of the fluid is constant, the presence of chemical potential gradients can still cause a pressure gradients at an interface. 

If we hold the pressure in the bulk constant ($\grad P=0$), 
we can express the solvent chemical potential ($\grad\mu_f)_{P, T}$ using Gibbs-Duhem in the \textit{bulk} as,
\eq{\label{eq:connection}(\grad \mu_f)_{P,T}=- \sum_{i=1}^{n-1} \frac{c_i^B}{c_f^B}(\grad \mu_i)_{P,T}\;.}
The dissipation function depends on $n-1$ chemical-potential gradients, plus the term due to an explicit pressure gradient:
\eq{\label{eq:entropy_production}\Phi= \mathbf{Q} \cdot (-\grad P) + \sum_{i=1}^{n-1}
\left(\mathbf{J}_i-\frac{c_i^B}{c_f^B}\mathbf{J}_f\right) \cdot (-\grad \mu_i)_{P,T}\;.}
In what follows, we focus on a two-component system, with solvent $f$ and solute $s$. 
The dissipation function then becomes
\eq{\label{eq:entropy_binary}\Phi= \mathbf{Q} \cdot (-\grad P) +
 \mathbf{J}'_s\cdot (-\grad \mu_i)_{P,T}\;,}
where we have defined the excess flux of solute as  
\eq{\mathbf{J}'_s =\mathbf{J}_s-\frac{c_s^B}{c_f^B}\mathbf{J}_f\;.}
Finally, we can write the transport matrix connecting the fluxes with the thermodynamic forces,

\eq{
\label{eq:Transport}
\begin{bmatrix}
    \mathbf{Q} \\
    \mathbf{J}'_{s}
\end{bmatrix}
=
\begin{bmatrix}
M_{QQ} & M_{QJ} \\
M_{JQ} & M_{JJ} \\
\end{bmatrix}
\begin{bmatrix}
    -\grad P/T \\
    -\grad \mu_s/T
\end{bmatrix}
}
By including the factor $1/T$  in the thermodynamic forces, we can cast the entropy production in a simple bi-linear form in fluxes and thermodynamic forces. Such form is needed to derive the Onsager reciprocity relations for the transport coefficients $M_{\alpha\beta}$. In practice, the factor $1/T$ is often absorbed in the transport coefficients.

\subsection{Transport coefficients}
\label{sec:GK}
To compute the transport coefficients $M_{\alpha\delta}$ in Eq. \eqref{eq:Transport} using FD-NEMD, we need to represent the thermodynamic forces as fictitious mechanical forces incorporated in the Hamiltonian of the system and that can act on the particles in the fluid.
Here, we recapitulate the derivation by Yoshida{\em et al.}~\cite{Yoshida2017} to show that such an approach provides the mechanical route to Onsager's symmetry relations.

We consider a system with $N$ interacting particles satisfying Hamiltonian equations of motion:
\eq{\mathbf{\dot r}_i=\frac{\mathbf{p}_i}{m_i}\;,}
\eq{\mathbf{\dot p}_i=\mathbf{F}_i+\mathbf{F}_{ext}\;,}
where $\mathbf{F}_i$ is the force exerted  on particle $i$ by all the other particles, and $\mathbf{F}_{\text{ext}}$ is the mechanical equivalent of the thermodynamic force.

For the diffusio-osmotic case, we represent all chemical potential gradients by equivalent forces $\mathbf{F}_i^{\mu}$ on every particle of species $i$. 
To satisfy the condition of mechanical equilibrium in the bulk, the force $\mathbf{F}_{s}^{\mu}$ on the solute particles must be balanced by a force $\mathbf{F}_{f}^{\mu}$ on the solvent particles, such that:
\eq{\label{eq:bulk_eq}\mathbf{F}^B_{\textnormal{ext}}=0=\mathbf{F}_s^{\mu}N_s^B+(N^B-N_s^B)\mathbf{F}_{f}^{\mu}\;,}
where $N^B$ and $N_s^B$ are the total number of particles and the number of solute particles in the bulk. 
Eq.~\eqref{eq:bulk_eq} is the mechanical equivalent of the Gibbs-Duhem equation.

Expressing everything in terms of the external force on the solutes
\eq{\mathbf{F}^B_\textnormal{{ext}}=\left[N_s^B-\frac{N_s^B}{N^B-N_s^B}(N^B-N_s^B)\right]\mathbf{F}_s^{\mu}=0\;.}
The Hamiltonian coupling of the particles to the external driving forces is 

\eq{\label{eq:H_ext}\mathcal{H}_{\text{ext}}=\left[ \sideset{}{^B}\sum_{i \in{s}} \mathbf{x}_i -\frac{N_s^B}{N^B-N_s^B}\sideset{}{^B}\sum_{i \in{f}} \mathbf{x}_i\right]\cdot \mathbf{F}_s^{\mu}\;.}
It is worth pointing out that all the sums in Eq.~\eqref{eq:H_ext} are in the bulk $B$.
Next, we consider a system confined in a slit. The total volume of the fluid $\Omega$ includes an interfacial region. The previous expression is still valid, giving rise to the diffusio-osmotic flow, as now there is a non-vanishing contribution from the externally applied forces $\mathbf{F}_\textnormal{{ext}}$

\eq{\mathcal{H}_{\text{ext}}=\left[ \sideset{}{^\Omega}\sum_{i\in{s}} \mathbf{x}_i -\frac{N_s^{B}}{N^{B}-N_s^{B}}\sideset{}{^\Omega}\sum_{i\in{f}} \mathbf{x}_i\right]\cdot \mathbf{F}_s^{\mu}\;.}
From linear response theory \cite{Hansen2006}, we can compute the response of a given observable $B$ to an external perturbation of the form $\Delta \mathcal{H}=A(x_i)F_{0}=\mathcal{H}_{\text{ext}}$ as
\eq{\label{GK} <B>=L_{AB}F_0=\left[ \frac{1}{k_BT} \int_0^{\infty}<B(t) \dot A(0)> dt\right]F_0\;.}
Focusing on the non-diagonal terms of the transport matrix on Eq.~\eqref{eq:Transport}, when a chemical potential gradient is applied, the observable we want to measure is the total flux of the particles $\mathbf{Q}$

\eq{\label{eq:Total_vol_flow}B=\mathbf{Q}^{\Omega}=\frac{1}{N^{\Omega}}\sideset{}{^\Omega}\sum_{i  \in {\text{all}}} \dot {\mathbf{x}}_i\;.}
It is convenient to write the variable that couples to the external field as
\begin{align*}
\dot A &= \sideset{}{^\Omega}\sum_{i \in{s}} \dot {\mathbf{x}}_i -\frac{N_s^{B}}{N^{B}-N_s^{B}}\sideset{}{^\Omega}\sum_{i \in{f}} \dot {\mathbf{x}}_i
\\
&=V^{\Omega}\left(\frac{1}{V^{\Omega}}\sideset{}{^\Omega}\sum_{i\in{s}} \dot {\mathbf{x}}_i -\frac{c_s^{B}}{c_f^{B}}\frac{1}{V^{\Omega}}\sideset{}{^\Omega}\sum_{i\in{f}} \dot {\mathbf{x}}_i\right) 
\\
&=V^{\Omega}\left( \mathbf{J}_{s}^{\Omega}-\frac{c_s^{B}}{c_f^{B}}\mathbf{J}_{f}^{\Omega}\right) \;.
\end{align*}
Finally, using Eq.~\eqref{GK} we can express the total volume flux as

\eq{\label{eq:Q_GK}\!\begin{aligned} \mathbf{Q}^{\Omega} = <\mathbf{Q}^{\Omega}> &= \left[ \frac{V^{\Omega}}{k_BT} \int_0^{\infty}<\mathbf{Q}^{\Omega}(t)  ( \mathbf{J}_{s}^{\Omega}-\frac{c_s^{B}}{c_f^{B}}\mathbf{J}_{f}^{\Omega})(0)> dt\right]\mathbf{F}_s^{\mu}
\\
&= M_{QJ}\frac{\mathbf{F}_s^{\mu}}{T}\;.
\end{aligned}
}
Hence, using transport equations in Eq.\eqref{eq:Transport}, we can establish the connection between the thermodynamic force and its microscopic counterpart as 

\eq{\label{eq:thermo_micro}\grad \mu_s =-\mathbf{F}_s^{\mu}\;.} 
Eq.~\eqref{eq:thermo_micro} is general (i.e. it is valid for arbitrary forces). 
However, the Green-Kubo expression in Eq.~\eqref{GK} is only valid in the linear regime, in which case the fluxes that appear in the expression for the entropy production  (Eq.~\eqref{eq:Transport}) are linear functions of the thermodynamic forces. 
Eq.~\eqref{eq:thermo_micro} seems to differ from the result reported by Yoshida {\em et al.}~\cite{Yoshida2017}, but this is only apparent: the discrepancy is due to an unfortunate definition for $\grad \mu_s$ in ref.~\cite{Yoshida2017}, which  is only correct in the limit of infinite dilution. 
As a consequence, $\mathbf{F}_s^{\mu}$ of ref.~\cite{Yoshida2017} is underestimated by a factor $\phi^B_f\equiv N_f^B/N^B$. 

We now focus on the off-diagonal term $M_{JQ}$ of the transport matrix. 
This coefficient expresses the dependence of the excess solute flux on the bulk pressure gradient. 
A pressure gradient exerts a force on a volume of fluid rather than on individual particles. 
As a first approximation, one might tend to connect the thermodynamic force acting on the system to the microscopic force as (see e.g \cite{Todd1995,Fu2017,Yoshida2017,Todd2017,Mangaud2020}):
\begin{equation}\label{eq:FgradP}
\mathbf{F}^{P} = -\grad P/ c^{\Omega}\;.
\end{equation}
 It is important to realize that in confined geometries, and {\em a fortiori} in porous media, it may be problematic to work with local pressure gradients even though it is perfectly legitimate to consider the pressure difference between the reservoirs on either side of the system.
The reason is that if the potential energies inside and outside the slit are different, $\grad P$ would show $\delta$-function spikes at the entrance and exit of the slit, whereas local thermodynamic equilibrium requires that all $\mu_i$s are continuous. 
If the properties of the slit are constant in the direction of the flux, the chemical potential gradients, and hence the color forces,  are constant inside the slit.  

Of course, due to interactions with the wall, the fluid density may vary in the direction perpendicular to the wall. 
In that case, a constant force per particle creates different pressure gradients at different distances from the wall.
This is not in contradiction with the statement that the pressures are the same everywhere  inside the reservoirs, precisely because the local pressure may vary rapidly at the entrance and exit of the channel.

In what follows, we consider a small volume $\omega$ at a distance $z_{\omega}$ from the wall. We obtain that the Hamiltonian coupling to the external force is

\eq{\mathcal{H}_{ext} = \sideset{}{^\omega}\sum_{i\in \textnormal{all}} \mathbf{x}_i \cdot \mathbf{F}^{P}\;,}
therefore $A(\omega) =\sideset{}{^\omega}\sum_{i\in \textnormal{s}} \mathbf{x}_i$.
The variable that couples to the external field $\mathbf{F}^{P}$ is given by
\begin{align*}
\dot A(\omega) &= \sideset{}{^\omega}\sum_{i\in \textnormal{s}} \dot{\mathbf{x}}_i
\\
&=V^\omega\left(\frac{1}{N^{\omega}}\sideset{}{^\omega}\sum_{i\in \textnormal{s}} \dot{\mathbf{x}}_i \right)\frac{N^\omega}{V^\omega}
\\
&=V^\omega\mathbf{Q}^{\omega}c(\omega)\;.
\\
\end{align*}
Finally, using Eq.~\eqref{GK} we can express the excess solute flux as:

\begin{align}
\label{eq:excess-solute}
 \mathbf{J}^{\Omega}_s-\frac{c_s^B}{c_f^B}\mathbf{J}^{\Omega}_f = <\mathbf{J}^{\Omega}_s-\frac{c_s^B}{c_f^B}\mathbf{J}^{\Omega}_f> &= \left[ \frac{V^{\omega}}{k_BT} \int_0^{\infty}<(\mathbf{J}^{\Omega}_s-\frac{c_s^B}{c_f^B}\mathbf{J}^{\Omega}_f)(t)  ({Q}^{\Omega})(0)> dt\right]c\,\mathbf{F}^{P}
\nonumber \\
&= \frac{M_{JQ}}{T}\,c\,\mathbf{F}^{P} \;. 
\end{align}
By comparing Eq.~\eqref{eq:excess-solute} with Eq.~\eqref{eq:Transport}   the pressure gradient that corresponds to a constant force per particle 
is given by:
\eq{\label{eq:gradP_force}\grad P(z) = - c(z)\mathbf{F}^{P} \;.}
We thus conclude that the expressions for the transport coefficients in Eq.~\eqref{eq:Q_GK} and Eq.~\eqref{eq:excess-solute} are equivalent, as the correlation functions are symmetric in time. 
Thus, $M_{JQ} = M_{QJ}$, fulfilling Onsager's reciprocal relations.This result suggests that to obtain the cross-coefficients, in principle, we can apply pressure gradients or chemical potential gradients. In practice, the advantage of the latter is that they do not depend on the distance with the interface.

\subsection{Local and global fluxes}
It is instructive to look at the expression for the entropy production in a system between two reservoirs at different chemical potentials. We will assume that the temperature of the system is constant.
In that case, the pressure in both reservoirs is a function of the chemical potentials and is therefore not an independent thermodynamic variable. 

The dissipation function for a macroscopic volume with  chemical-potential profiles  $\mu_i({\bf r})$, where  $i$ labels the $n$ different species, is given by:
\begin{equation}\label{Phi_DO}
\Phi = \oint_S d{\bf S}\cdot\left(
 \sum_{i=1}^n\mu_i({\bf r}) {\bf j}_i({\bf r})  \right)\; ,
\end{equation}
where  the ${\bf j}_i({\bf r}) $ denote the diffusive fluxes of species $i$, and the integral is over  the surface of the system.
We  focus on the practically important case that the system is in contact with two external reservoirs (denoted by $I$ and $II$) that, individually, are at constant  $\mu_i$. These reservoirs are not in direct contact with each other. 
In that case, the boundary conditions are completely specified by the  $\mu_i^I$ and $\mu_i^{II}$. The global dissipation function of the system is then given by 
\begin{equation}\label{eq:sigma2}
\Phi= \sum_{i=1}^n \left(\mu_i^{II}- \mu_i^{I} \right)J_i =
\sum_{i=1}^n \frac{\left(\mu_i^{II}- \mu_i^{I} \right)}{L} j^x_i S
\; ,
\end{equation}
where  $J_i$ denotes the total flow of particles of species $i$  from $I$ to $II$, {\em i.e.}, the total number of particle of species $i$ that crosses either surface per unit time. 

We  can use Gauss's theorem to rewrite Eq.~\ref{eq:sigma2} as
\begin{equation}\label{eq:sigma3}
\Phi= \int_V d{\bf r} \; 
\sum_{i=1}^n {\bf\nabla }(\mu_i({\bf r})\cdot {\bf j}_i({\bf r})) \; .
\end{equation}
We can rewrite this as:
\begin{eqnarray} \label{eq:sigma4}
\Phi &=& \int_V d{\bf r} \; \left[ \sum_{i=1}^n {\bf\nabla }\mu_i({\bf r}) \cdot{\bf j}_i({\bf r})+
\sum_{i=1}^n \mu_i({\bf r}) {\bf\nabla }\cdot{\bf j}_i({\bf r}) \right]\;.
\end{eqnarray}
We note that, in steady state, the divergences of all fluxes must vanish. Hence the second line of Eq.~\ref{eq:sigma4} vanishes and we are left with
\begin{equation}
\Phi= \int_V d{\bf r} \; 
 \sum_{i=1}^n {\bf\nabla }\mu_i({\bf r})\cdot{\bf j}_i({\bf r}) 
\end{equation}
Note that adding the rotation of a vector field to the fluxes will not change this result, provided that the normal component of this rotation at the boundaries vanishes. 
Another way of saying the same thing is that Eq.~\ref{Phi_DO} shows that adding any flux ${\bf j}'$ that vanishes at the boundaries of the system (or, at least, is purely tangential to the boundaries), will not contribute to the entropy production.
The above argument also holds for other fluxes, such as the heat flux, which,  in contrast to the heat flow into and out of a system, are not uniquely defined.  

\subsection{Local Thermodynamic Equilibrium and the Derjaguin-Anderson theory for diffusio-osmosis}
\label{sec:Derjaguin-Anderson}
We consider again the system in Fig.~\ref{fig:DO_system}. The mixture is at a constant temperature and, we assume a chemical potential gradient of species $i$ in the $x$-direction. If the bulk fluid is incompressible, hence, the density and pressure equilibrate instantaneously in this region. 
Moreover, the rate of the spontaneous decay of chemical potential gradients over a distance $\ell$ scales as $\ell^2/D_i$ ($D_i$ denotes the diffusion coefficient of species $i$). As a consequence, chemical potential differences across the boundary layers equilibrate very quickly compared to the time scale of the diffusio-osmotic flow. Therefore, we can employ local thermodynamic equilibrium, assuming that the system is in equilibrium in the $z$-direction, even though a chemical potential gradient can be maintained along the $x$-direction. Hence, we can write the relation between the thermodynamic forces in the bulk from Eq.\eqref{eq:GD_connection} as:

\eq{\pdv{P_{xx}^B}{x}=0=\sum_{i=1}^n c_i^B\left(\pdv{\mu_i}{x}\right)\;,}
where $P_{xx}$ refers to a component of the pressure tensor parallel to the surface.
At the interface, the density profile $c_i(z)$ depends on $z$. The fact that $\mu$ = $\mu^{\textnormal{exc}}(z)$ + $k_BT\ln c_i(z)$ is constant across the diffusive boundary layer (and for a fixed $x$) implies that the excess chemical potential $\mu^{\textnormal{exc}}$ will, in general, depend on the distance $z$ from the wall. 
At a point $z$ within the diffusive boundary layer we can write,

\eq{\label{PressureGrad}\pdv{P_{xx}(z)}{x}=\sum_{i=1}^n [c_i(z)-c_i^B]\left(\pdv{\mu_i}{x}\right)\;.}
Once more, it is important to stress that mechanical forces in liquids can only be caused by body forces such as gravity or by pressure gradients \cite{Ganti2017}. 
The reason why chemical potential gradients near a surface cause fluid flow is that they induce a pressure gradient near a wall. It is the pressure gradient in Eq.~\eqref{PressureGrad} which moves the fluid. 

As the chemical potential $\mu$ is constant, we can relate the concentrations in the bulk ($z\to \infty$) and close to the surface as
 
\eq{c_i(z)e^{\beta (\mu_i^{\text{exc}}(z))} =c_i^B e^{\beta (\mu_i^{\text{exc}}(\infty))}\;,} 
thus, we can rewrite Eq.\eqref{PressureGrad},
\eq{\label{eq:grad_p}\pdv{P_{xx}(z)}{x}=\sum_{i=1}^n c_i^B[e^{-\beta \Delta \mu_i^{\text{exc}}(z)}-1]\left(\pdv{\mu_i}{x}\right)\;.}
where $\Delta \mu_i(z)^{\text{exc}} = \mu_i^{\text{exc}}(z)-
\mu_i^{\text{exc}}(\infty)$ is the excess chemical potential due to the presence of the interface. 
We can now combine Eq.~\eqref{eq:grad_p} with the Stokes equation to estimate the flow velocity in the $x$ direction:

\eq{\eta(z) \frac{\partial^2 v_x(z)}{\partial z^2}=\pdv{P_{xx}(z)}{x}\label{Sx}\;.}
Assuming a constant viscosity $\eta$,we get

\eq{\label{eq:v_LTE}v_x(z)=-\frac{1}{\eta}\int_0^z dz'\int_{z'}^\infty dz''\sum_{i=1}^n c_i^B[e^{-\beta \Delta \mu_i^{\text{exc}}(z)}-1]\left(\pdv{\mu_i}{x}\right)\;.}
We note an important feature of Eq.~\ref{eq:v_LTE}: it is {\em not} expressed in terms of the pressure gradients, which are microscopically ill-defined in an inhomogeneous system (e.g near a surface). This in contrast a chemical potential gradients which, as we know, can be replaced by a uniquely defined force per particle.
The other point to note is that we have {\em assumed} that the macroscopic creeping-flow approximation holds. This assumption is, in general, not correct, and in simulations we do not make this assumption. 

Using non-slip boundary conditions, and exploiting the fact that outside the diffuse layer, the  velocity does not vary, we obtain the bulk velocity of the fluid $v_B$,

\eq{\label{eq:vb2_LTE}v_x(z\to\infty)=v_x^B=-\frac{1} {\eta}\int_0^\infty dz \,z \sum_{i=1}^n c_i^B[e^{-\beta \Delta \mu_i^{\text{exc}}(z)}-1]\left(\pdv{\mu_i}{x}\right)}
Note that in fluid dynamics, the slip velocity is usually defined as the velocity at the interface where the boundary condition is imposed. However, in the present case, using a local continuum description, the  slip velocity $v_x^B$ is the fluid velocity in the bulk just outside the diffuse layer. 

The Derjaguin-Anderson description of diffusio-osmosis~\cite{Derjaguin1947,Anderson1989} can be obtained as a special case of Eq.~\eqref{eq:vb2_LTE}.
For an ideal bulk solution, we have:

\eq{\label{eq:gradc_grad_mu}\pdv{\mu_i}{x}= \frac{k_BT}{c_i^B} \pdv{c_i^B}{x}\;.}
Thus, we can write Eq.~\eqref{eq:vb2_LTE} as,

\eq{v_x^B=-\frac{k_BT} {\eta}\int_0^\infty dz \,z \sum_{i=1}^n [e^{-\beta \Delta \mu_i^{\text{exc}}(z)}-1]\left(\pdv{c^B_i}{x}\right)\;.}
If we now restrict the analysis to very dilute solutions of  solute molecules $s$ in a continuum liquid phase (solvent $f$), 

\eq{\label{eq:v_Derjaguin-Anderson}v_x^B\approx -\frac{k_BT} {\eta}\int_0^\infty dz \,z [e^{-\beta \phi(z)}-1]\left(\pdv{c^B_s}{x}\right)\;,}
where we neglected the solvent contribution as $e^{-\beta \Delta\mu_i^{\text{exc}}(z)}\propto 1/c_i^B $ and $c_s^B \ll c_f^B$. Additionally, we defined $\phi(z)\equiv \Delta \mu_i^{\text{exc}}(z)$. In Derjaguin-Anderson theory $\phi(z)$ is the mean-field potential acting on  solutes at a distance $z$ from the solid surface. This potential does not only include the direct effect of the surface on the solutes, but accounts for the perturbation of the local liquid structure near a wall, as illustrated in Fig.~\ref{fig:DO_system}. 
We note that using the excess chemical potential instead of $\phi$ has the advantage that the expression $c_i(z)/c_i^B$= $[e^{-\beta \Delta \mu_i^{\text{exc}}(z)}-1]$ follows directly from the fact the chemical potential depends only on $x$ and not on $z$. Hence we can write 
\[
\mu_i(c)=k_BT\ln c_i(z)+\Delta\mu^{exc}_i(z) = k_BT\ln c_i^B\;
\]
To quantify whether the net effect of this layering is an accumulation or depletion of particles near the surface, we use Gibbs's definition of the surface excess for particles of species $i$ as (see~\cite{Anderson1984}):
\eq{\label{eq:gamma}\Gamma_i=\int_0^\infty[c_i(z)-c_i^B] dz\;.}
$\Gamma_i$ is positive if there is net adsorption of particles on the wall, and negative in the case of depletion at the interface. 
Following Anderson, we define  the so-called solute {\em adsorption-length} $K_i$ to the zeroth moment of the excess-concentration profile:
\eq{\label{eq:K}K_i=\int_0^\infty[c_i(z)/c_i^B-1] dz\;.}
$K_i$ can be interpreted as the thickness (positive or negative) of a layer of bulk solution that would contain the same net number of adsorbed or depleted particles. 
$K_i$ is obtained experimentally by equilibrium adsorption studies and it can be as large as 1 $\mu m$~\cite{Anderson1984}, or even much larger near a wetting transition. 

A second measure of the adsorption/depletion layer is given by  $\xi_i$. $\xi_i$, which has the dimensions of length squared,  is related to the first moment of the excess concentration:

\eq{\label{eq:xi}\xi_i\equiv\int_0^\infty[c_i(z)/c_i^B-1] z\, dz\;.}
Derjaguin defined the characteristic extension of the diffuse adsorption layer as $\sqrt{\xi_i}$ \cite{Derjaguin1987}, and Anderson defined the characteristic length $L_i^*$~\footnote{Anderson's definition can give unphysical results when $K_i\rightarrow 0$, while $\xi_i\ne$0.}:
\eq{\label{eq:L}L_i^*\equiv\frac{\xi_i}{K_i} \;.}
Using the above definitions, we can rewrite the diffusio-osmotic velocity in Eq.~\eqref{eq:v_Derjaguin-Anderson}
\eq{\label{eq:flatplate_do} v^B=-\frac{\alpha }{\beta \eta} K_sL_s^*\;,}
where $\alpha$ is the concentration gradient of solutes in the bulk.
Note that even when there is strong net adsorption of solutes (large $K_s$), $L_s^*$ may be small, zero, or even of the opposite sign, depending on  $(c_s(z)-c_s^B)$.  In other words, diffusio-osmotic flow is less sensitive to the excess concentration closest to the wall.
This effect becomes  very pronounced for thick adsorption layers, in particular near a wetting transition.

\subsection{Simulations}
\label{sec:ch_do_simulations}

\begin{figure}[H]
\centering
\includegraphics[width=0.8\linewidth]{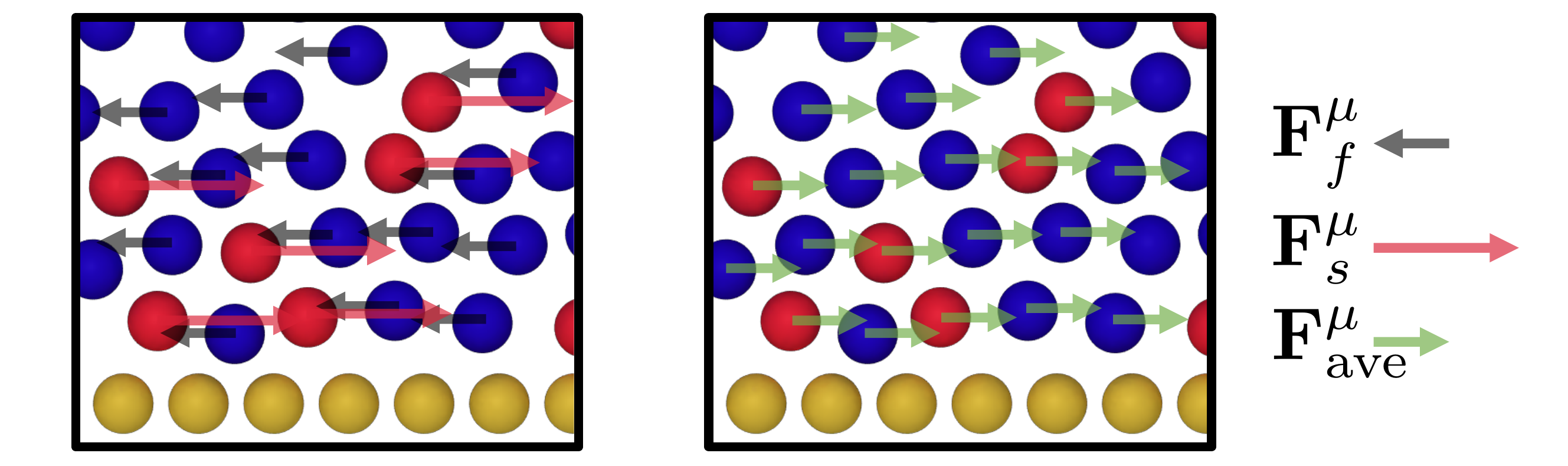}
\caption{Different methods to impose microscopic forces for diffusio-omotic(phoretic) simulations. On the right-hand side, a force $\mathbf{F}^{\mu}_i$ is applied on each particle depending on their species $i$. On the left-hand side, the average force for each particle $\mathbf{F}^{\mu}_{\text{ave}}$ computed using Eq~\eqref{eq:AverageForce_mu} is shown.}
\label{fig:Force_per_volume}
\end{figure}

There are two ways of imposing microscopic forces for diffusio-osmosis using FD-NEMD~\ref{fig:Force_per_volume}. As discussed above,  we can mimic a chemical potential gradient by applying color forces  on all particles~\cite{Yoshida2017}.
Alternatively,  and more in the spirit of hydrodynamics, we can start from  the force per volume element~\cite{Liu2018}, and then express the force per particle as
\eq{\label{eq:AverageForce_mu}\mathbf{F}^{\mu}_{\text{ave}}(z)=\frac{[c_s(z)\mathbf{F}^{\mu}_s+c_f(z)\mathbf{F}^{\mu}_f]}{c(z)}}
where the force on the solutes $\vb{F}^{\mu}_s$ is given by Eq.~\eqref{eq:thermo_micro} and the force on the solvents $\vb{F}^{\mu}_f$ is determined by imposing mechanical equilibrium in the bulk (see Eq. \eqref{eq:bulk_eq}). Both approaches should give the same flow profiles if the spatial binning used to measure the concentration distributions in Eq.~\eqref{eq:AverageForce_mu} is the same as the one used to sample the velocity profiles. 

In practice there is a difference: the force $\mathbf{F}^{\mu}_i$ on solute and solvent particles is the same throughout the system.
However, the net force per volume element is non-zero only close to the wall.
In our simulations, we make use of this fact by imposing $\mathbf{F}^{\mu}_{\text{ave}}(z)$ =0 at distances that are sufficiently far away from the wall for the density modulations to have decayed. 
The advantage of this approach is the following: due to the dynamic adsorption/desorption of solute and solvent particles near the wall, there will be small -- in the thermodynamic limit: vanishingly small -- composition fluctuations in the bulk. 
However, even though these fluctuations are small, their integrated effect is non-negligible: they would result in a flow in the  direction opposite to the interface-induced flow. 
This spurious bulk flow is a finite-size effect, in the sense that, for a sufficiently large wall area, positive and negative density fluctuations will cancel.
However, in a finite system, we need to suppress this spurious bulk flow explicitly.
This we do by truncating the force per unit volume outside the interfacial region.
In Sec.~\ref{sec:Colloid} we will consider a more complex geometry where we cannot easily work with the average force, and we will describe an alternative method to suppress the spurious bulk flow.

We benchmarked our simulations against the published results of ref.~\cite{Yoshida2017}.
All simulations were perfomed using the  LAMMPS software package~\cite{Plimpton1995}. Particles interact via a 12-6 Lennard-Jones potential (LJ) $V_{LJ}(r)=4\epsilon^{LJ}_{ij}[(\sigma^{LJ}_{ij}/r)^{12}-(\sigma^{LJ}_{ij}/r)^6]$ shifted and truncated at $r= r_{\text{cut}}$, such that 
\begin{equation}
    V_{TS}(r)=
    \begin{cases}
      V_{LJ}(r)-V_{LJ}(r_{\text{cut}}), & \text{if}\ r \le r_{\text{cut}} \\
      0, & \text{otherwise}.
    \end{cases}
    \label{eq:LJ}
  \end{equation}
The indices $i$ and $j$ denote the particle types in our simulations: solutes ($s$), solvents ($f$) and wall ($w$). 
We chose the same Lennard-Jones interaction for the particle pairs $ss$, $sf$, $ff$ with $\epsilon^{LJ}_{ij}=\epsilon_0$ and $\sigma^{LJ}_{ij}=\sigma_0$, such that the bulk solution is  an ideal mixture. 
For convenience (Ockham's razor) we also use these same parameters for the wall-solvent interaction $wf$. 
We also assume that all particles have equal mass.
The wall-solute interaction strength $\epsilon^{LJ}_{ws}$ and $\sigma^{LJ}_{ws}$ were varied to control the degree of solute adsorption or depletion. 
For all interactions, $r_c = 2.5 \sigma_0$. In what follows, we use the mass $m_0$ of all the particles ($s$,$f$ and $w$) as our unit of mass and 
we set our unit of energy equal to $\varepsilon_0$, whilst our unit of length is equal to $\sigma_0$. 
All other units are expressed in terms of these basic units. 

\begin{figure}[H]
\centering
\includegraphics[width=0.80\linewidth]{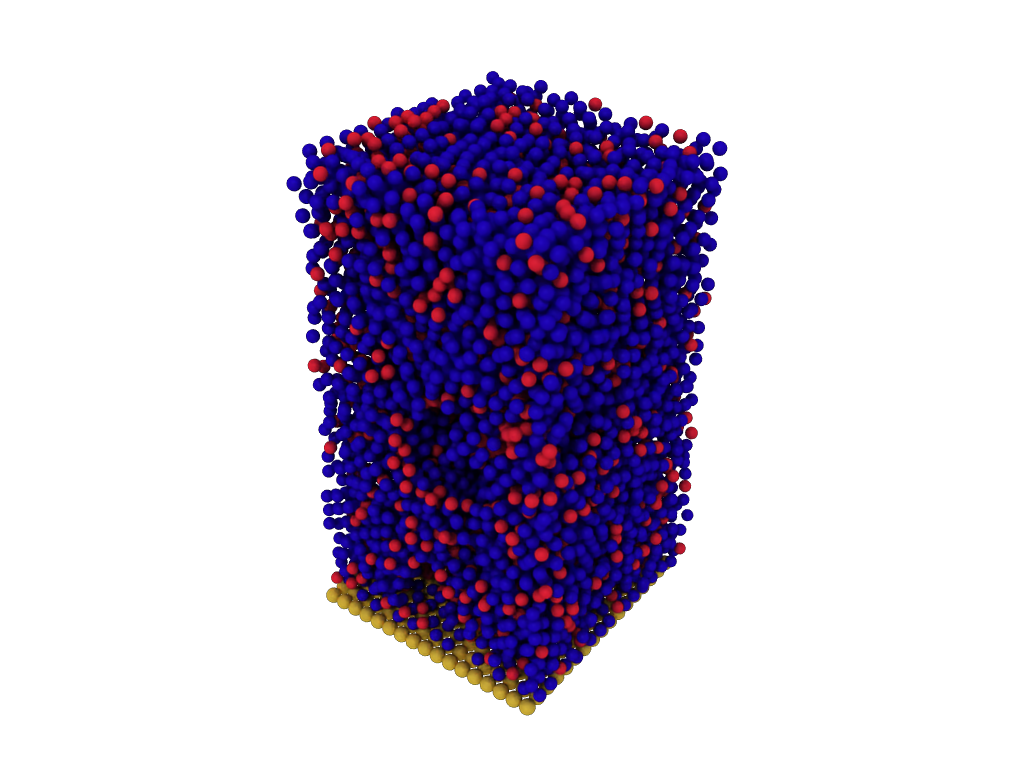}
\caption{Simulation box used for NEMD simulations. Periodic boundary conditions are used in the $x$ and $y$ directions. 
In the $z$ direction, the bottom wall has the structure of an FCC (100) surface whilst particles undergo specular reflection at the top surface.}  
\label{fig:Benchmark_system}
\end{figure}

A snapshot of the system in contact with the confining wall is shown in Fig.~\ref{fig:Benchmark_system}. 
The initial dimensions of the simulation box are $(17\sigma_0,17\sigma_0,35\sigma_0)$ with 7424 solution particles. 
The average concentration of solutes in the whole volume of $\bar c_s =0.15$.
The box is periodic in the $x$ and $y$ directions. In the $z$ direction, there is a solid wall at the bottom and a moving surface at the top, where particles undergo specular reflection. 
As the tangential momentum of particles remains unchanged upon reflection, this wall imposes ``slip'' boundary conditions. 
In contrast,  lower surface consists of a layer of immobile solid atoms with the structure of the (100) surface of  a face-centred cubic (FCC) lattice with lattice  constant $\sqrt{2}\sigma_0$. The interaction parameters of the solutes with the wall are $(\varepsilon_{sw},\sigma^{LJ}_{sw})=(1.5,1.5)$. 
We used a Nos\'{e}-Hoover thermostat~\cite{Hoover1985} to fix $k_BT/\varepsilon_0=1.0$ for all the simulations.

To initialise the system, we performed $10^5$ NVT MD steps, using a time step $\Delta t=0.002 \tau$.  
$5\times10^5$  steps were required to impose $P\sigma_0^3/\varepsilon_0=1.0$ as described in \cite{Yoshida2014,Yoshida2016}; this was achieved by allowing the box height to fluctuate, with the imposed  pressure applied to the moving wall.
 During this process, we sampled the height in the $z$-direction. 
 For all the subsequent simulations, the height was fixed at the average value of this fluctuating height.

\begin{figure}[H]
\centering
\includegraphics[width=0.8\linewidth]{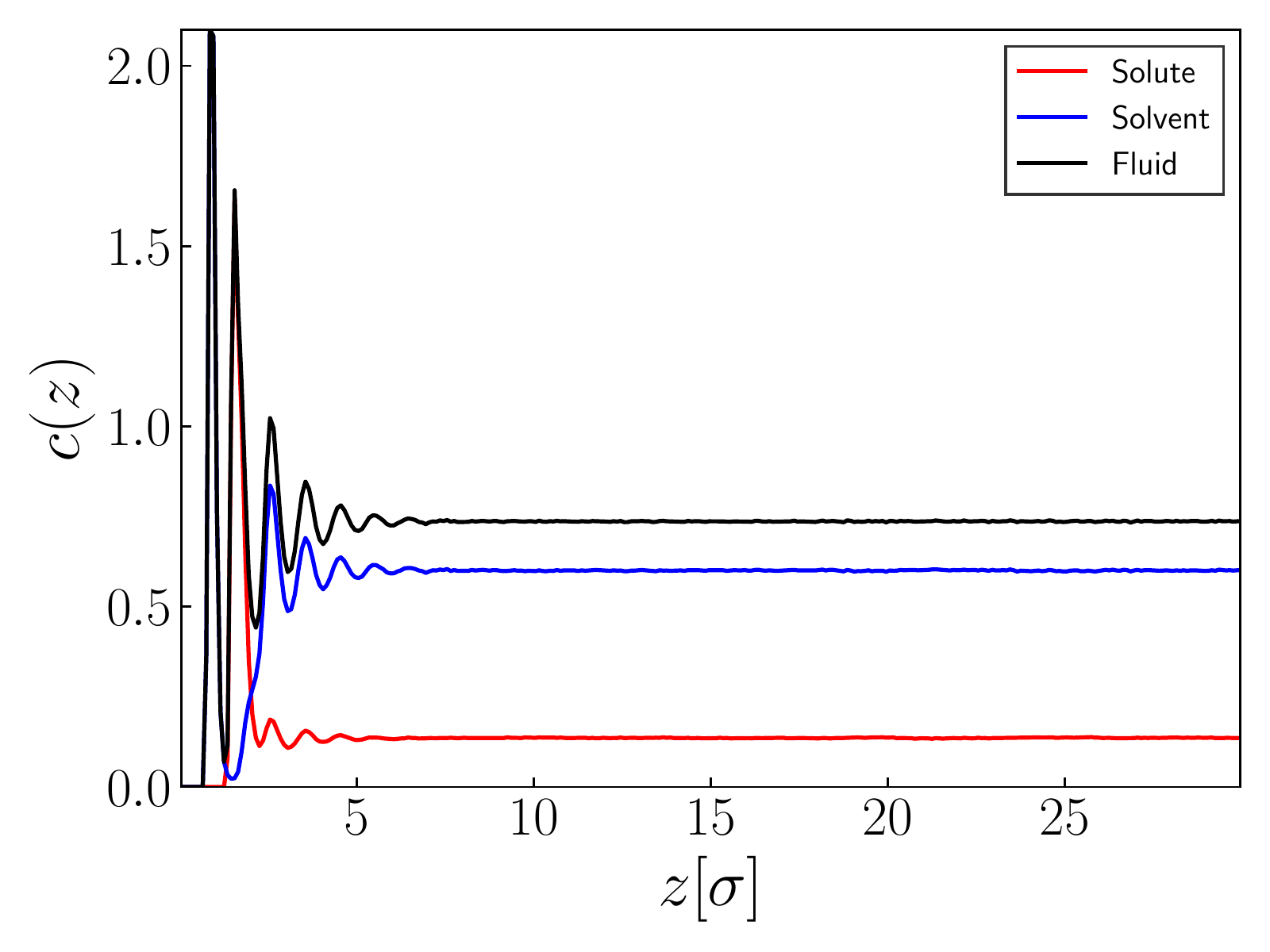}
\caption{Density distribution per species computed with a spatial resolution of $\Delta_z =0.25\sigma_0$. 
The imposed bulk concentration of the solutes was  $\bar c_s =0.15$. 
The first peak in the solvent density appears closer to the wall than that for the solutes because $\sigma_{wf}<\sigma_{ws}$.
}
\label{fig:concentration}
\end{figure}

After equilibrating the system, we sampled the density distribution for all the species during $3\times 10^6$ steps (see Fig.~\ref{fig:concentration}). The initial peak of the solvents near the wall is due to the fact that the wall-solute repulsion is stronger than the wall-solvent interaction.
 The migration of some solute particles towards the interface during the equilibration decreases their concentration in the bulk. 
 Therefore, $c_s^B < \bar c$. 
 However,  the effect is negligible for the system size and the relatively weak solvent adsorption, provided that we use the simulation technique described below.
As the reflecting top surface is hard, there is also some layering of the fluid there (see Fig.~\ref{fig:concentration}).
However, as there is no specific adsorption or depletion at the reflecting wall, it does not contribute to phoretic transport, and we can ignore it in our subsequent analysis.

Liu {\em et al.}~\cite{Liu2018} have shown that the flow profiles obtained using Eq.~\eqref{eq:AverageForce_mu} are in good agreement with results obtained applying an explicit chemical potential gradient. 
 As explained above, the application of $\mathbf{F}^{\mu}_{\text{ave}}(z)$ will not reproduce the correct diffusive fluxes, but that is not important in the context of this paper.
One disadvantage of Eq.~\eqref{eq:AverageForce_mu} for computations is that it requires knowledge of the equilibrium concentration profiles. 
In the present case, we have computed these profiles in a separate simulation, using a bin-width of 0.25 $\sigma_0$. 
However, it would probably be better to use the ``bin-less'' method of refs.~\cite{bor131,her181}. 

A conceptual disadvantage of working with $\mathbf{F}^{\mu}_{\text{ave}}(z)$ rather than the forces per species, is that $\nabla \cross \mathbf{F}^{\mu}_{\text{ave}}(z)$ need not be zero. 
In contrast, the rotation of the color forces vanishes. 

\begin{figure}[H]
\centering
\includegraphics[width=0.8\linewidth]{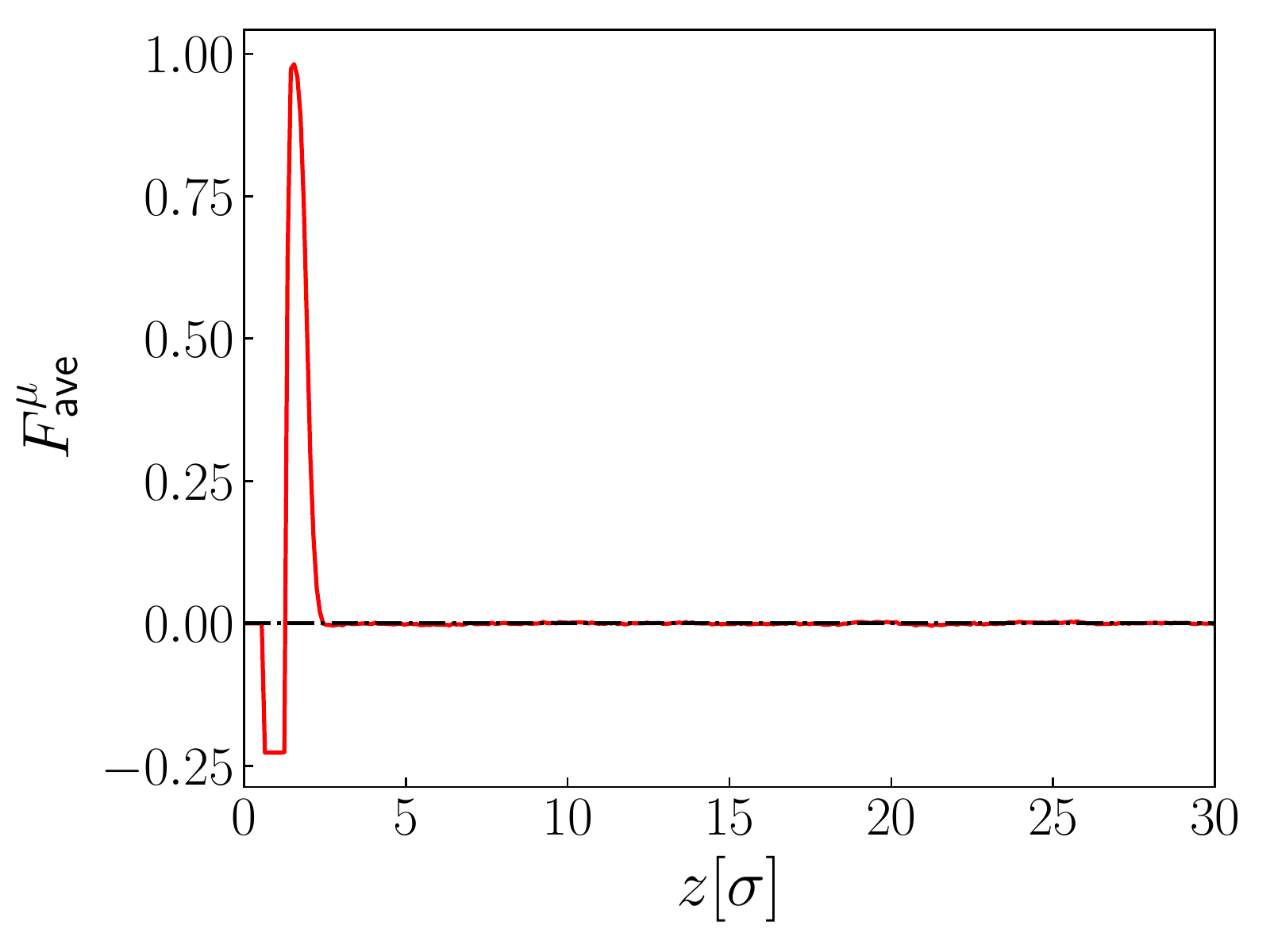}
\caption{Force applied as a function of the distance from the wall for diffusio-osmotic simulations with $\grad \mu_s = 1$. 
The force is shown in arbitrary units. Note that the net force per volume element vanishes for $z>4\sigma_0$, even though the color forces on the individual particles do not vanish. The imposed bulk concentration of the solutes was  $\bar c_s =0.15$.  }  
\label{fig:force_mu}
\end{figure}

The force distribution on the solution is shown in Fig.~\ref{fig:force_mu}.
From this figure it is clear that the net force per volume element is negligible for $z>4\sigma_0$. 

\begin{figure}[H]
\centering
\includegraphics[width=0.8\linewidth]{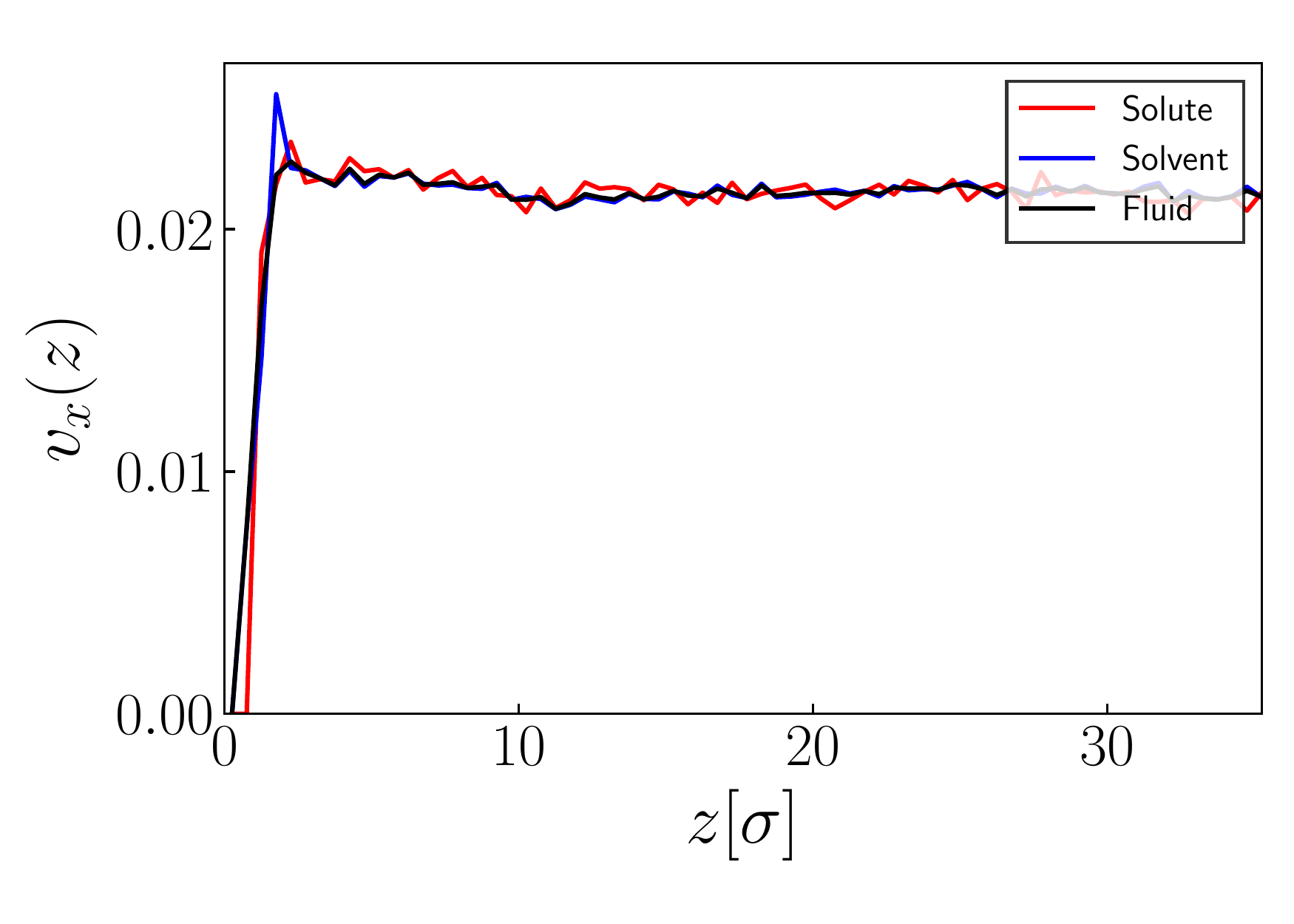}
\caption{Velocity profiles per species for diffusio-osmotic flow for $\grad \mu_s = -0.125$.  The bin size is $\Delta_z=0.25\sigma_0$.}  
\label{fig:do_flow}
\end{figure}

To obtain good statistics for the diffusion osmotic flows, we needed long simulations ($10^8$ time steps).
We applied the computed force distributions and measured the velocity profiles in the fluid. 
Results in Fig.~\ref{fig:do_flow} show the diffusio-osmotic velocity profile for $\grad \mu_s = -0.125$.We observe the plug-flow profile characteristic of diffusio-osmosis. 
At the interface, there is initially a steep increase in velocity due to the excess of solutes. Comparing with the benchmark, our results show a higher diffusio-osmotic velocity which comes from the fact that the color force used in ref~\cite{Yoshida2017} underestimates the effect of the thermodynamic force $\grad \mu_s$ by a factor equal to the molar fraction of solvents in the bulk $\phi^B_f=N_f^B/N^B$.
Notice that all the flow profiles are non-monotonic in $z$ and exhibit a peak before settling down to the bulk velocity. 
This peak has also been observed in previous studies~\cite{Yoshida2017,Liu2018}. This overshoot can only be partially be described using Eq.~\eqref{eq:v_LTE}. The remaining disagreement is not surprising, as Eq.~\eqref{eq:v_LTE} assumes that we can use the macroscopic creeping-flow equations with constant viscosity.

\subsection{Comparison with Theory}
\label{sec:Theo_predic_do}
In order to evaluate the theoretical expressions (Eq.~\eqref{eq:v_LTE} or Eq.~\eqref{eq:v_Derjaguin-Anderson}) for the slip velocity, we need to compute the concentration distribution $c_i(z)$ of all species $i$ as a function of the distance $z$ from the wall, and the viscosity $\eta$ of the solution. 
The former is obtained from EMD simulations, and relatively short runs are required as the equilibration in the $z$-direction is fast. 
In the simplest theoretical description, the viscosity is assumed to be independent of $z$, and equal to its bulk value: $\eta(z) = \eta^B$. $\eta^B$ was obtained using the Green-Kubo expression~\cite{Hansen2006}. 
Assuming that $\eta$ is independent of $z$ is a strong assumption, as we know that the fluid shows layering near the wall.

All relevant parameters in Sec.~\ref{sec:Derjaguin-Anderson}, such as $\Gamma$, $K$, $L^*$, depend on moments of the concentration distributions. 
The integrals in the definition of these parameters are evaluated from the surface ($z_0 =0$) to the bulk ($z \to \infty$). 
However, on a microscopic scale, the location of $z_0$ is problematic, the more so as the particle-wall interactions are different for solvent and solute as they have  different $\sigma^{LJ}$. 
\begin{figure}[H]
\centering
\includegraphics[width=0.8\linewidth]{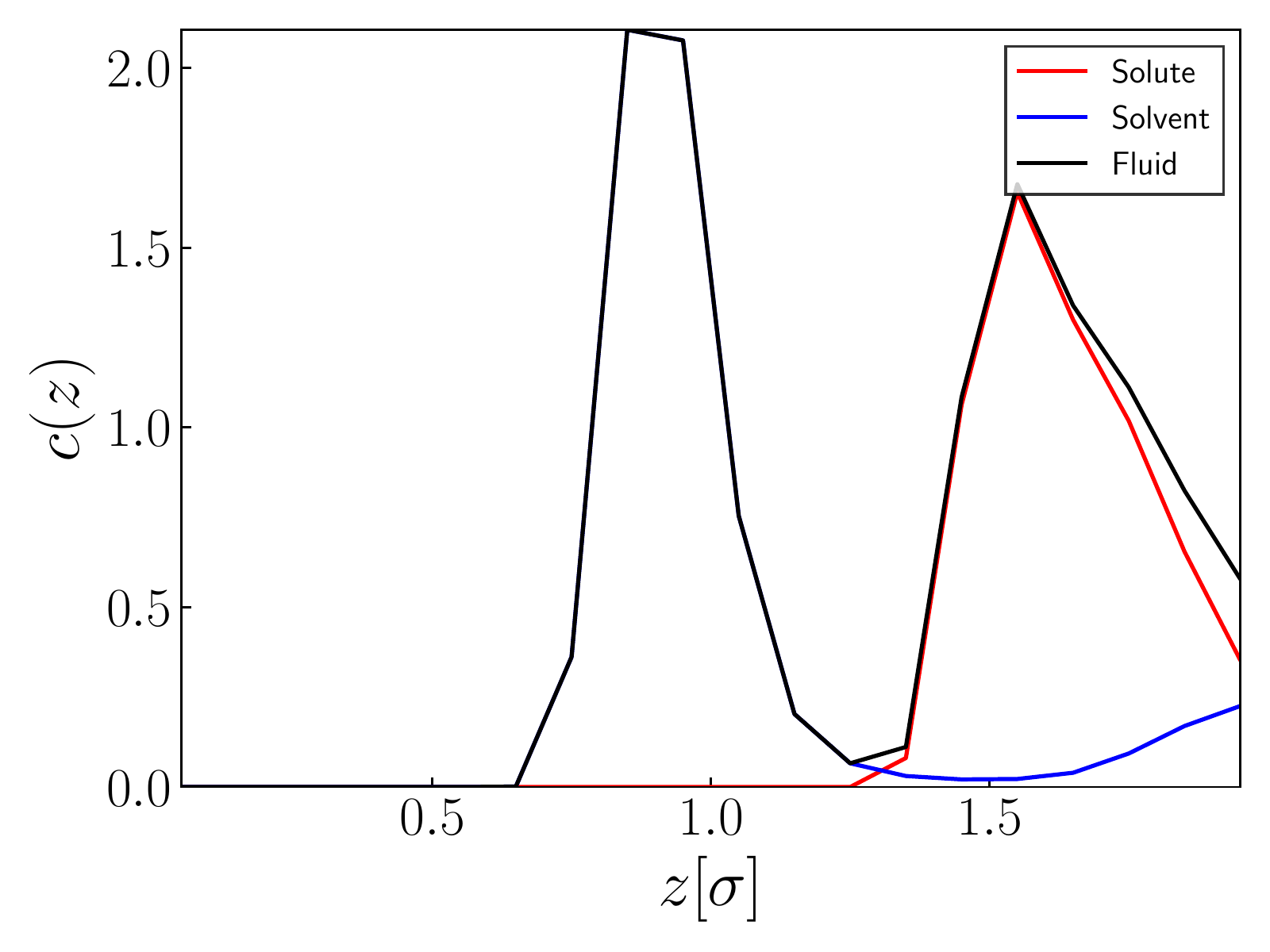}
\caption{Detail of the density profiles close to the solid wall. The bin size for sampling the density profile was $\Delta_z=0.1\sigma^{LJ}_0$.}  
\label{fig:zoom_rho_profile}
\end{figure} 
In Fig.~\ref{fig:zoom_rho_profile} we show the density profile for each species close to the wall. 
We can define a distance $d_i^{\text{min}}$ as the shortest distance to the wall where particles of species $i$ can penetrate. 
This distance is different for solvent and solute:
we obtain $d_f^{\text{min}}=0.55$ and $d_s^{\text{min}}=1.15$ for the solvents and solutes respectively. 
But the uncertainty in the location of the wall is not the only problem with the Derjaguin-Anderson theory:  if the adsorption of one or more species on the wall is very strong, we should expect the local viscosity  to become large and the strongly adsorbed layer will not contribute  to diffusio-osmosis. 

\begin{figure}[H]
\centering
\includegraphics[width=0.8\linewidth]{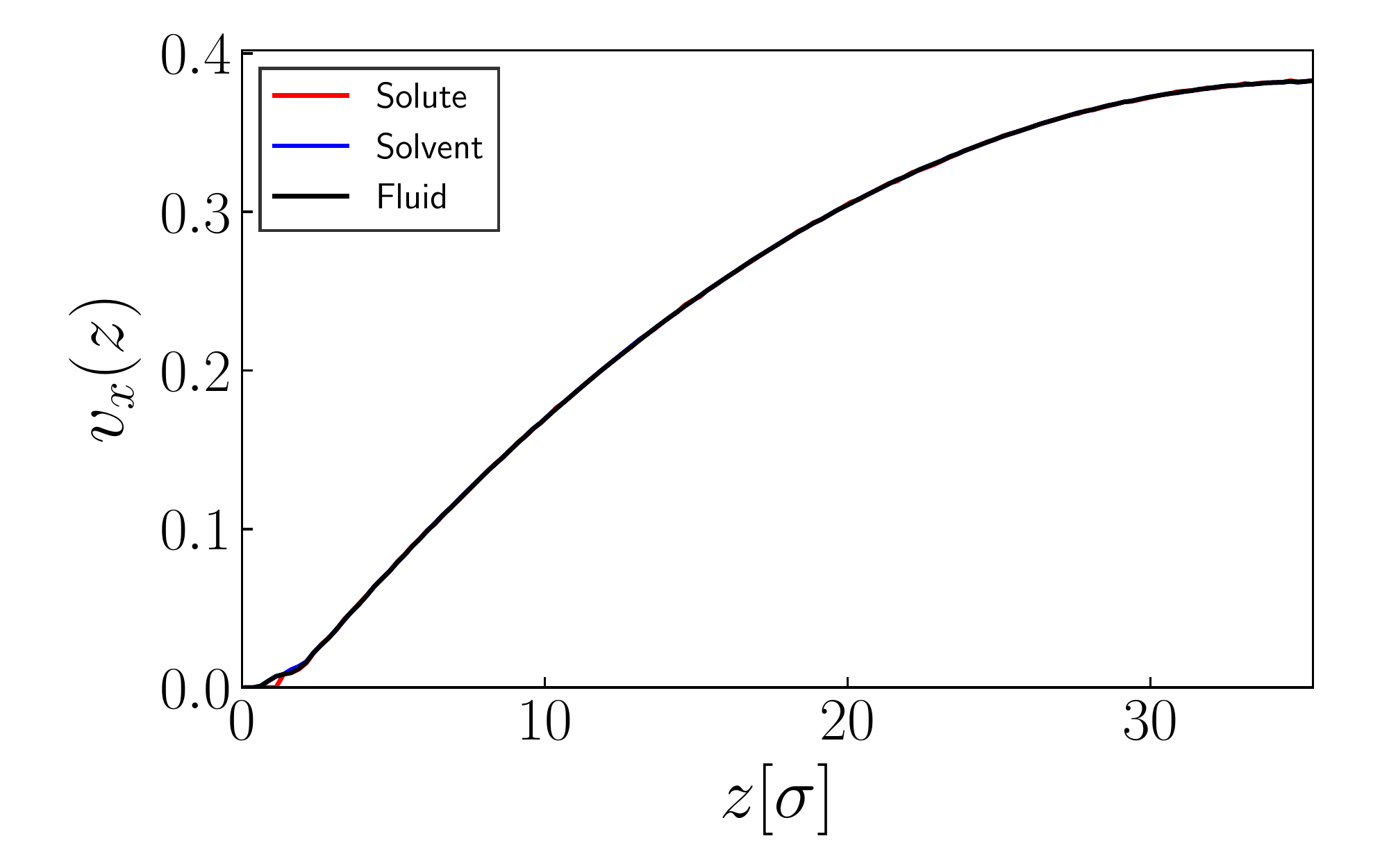}
\caption{Velocity profiles for solvent and solute, for the case of pressure-driven flow. The applied pressure gradient  in the bulk is  $\grad P = - 0.001$.
The flow profiles are sampled in bins of width $0.25\Delta_z=\sigma_0$.}  
\label{fig:p_driven_flow}
\end{figure}

This non-uniqueness of the location of the boundary in Eqn.~\ref{eq:xi} makes a comparison between theory and simulation difficult. 
In fact, there are two problems: a) the location of the boundary is different for solutes and solvents, but more importantly, b) a direct simulation of pressure-driven Poisseuille flow in the channel shows that the viscosity close to the wall is clearly higher than the bulk value, resulting in a smaller slope of the, otherwise parabolic flow profile close to the wall.
The latter effect can be seen in Fig.~\ref{fig:p_driven_flow}.


For phoretic transport, which depends on a the adsorption or depletion in a microscopic surface layer, the problems with the definition of $z_0$ are serious.

\begin{figure}[H]
\centering
\includegraphics[width=0.75\linewidth]{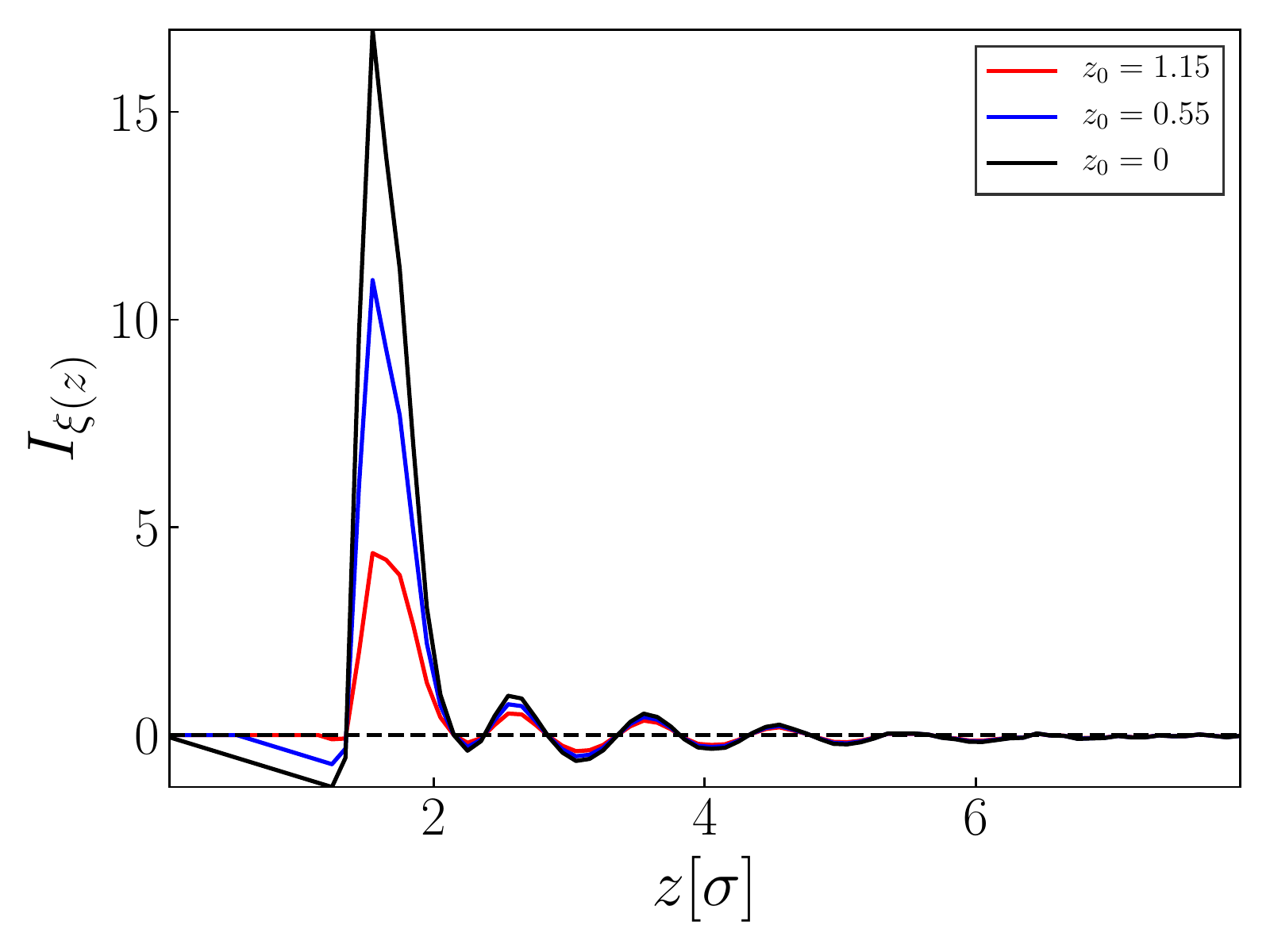}
\caption{Integrand in Eq.~\eqref{eq:xi} for the solutes with a concentration distribution sampled every $\Delta_z=0.25\sigma_0$. $z_0$ takes 3 different values: the position of the wall $z_0=0$, $z_0 =d_f^{\text{min}}=0.55$ and $z_0 =d_s^{\text{min}}=1.15$. Note that the integrands were translated accordingly such that their maximum values coincide.}  
\label{fig: L_start}
\end{figure}

We note that the most important parameter, $L^*$,  depends on the first moment of the concentration profile. 
Fig.~\ref{fig: L_start}  how strongly the integrand in Eq.~\eqref{eq:xi} depends on the assumed value of  $z_0$. 
The results show that in the case that we consider it is almost meaningless to attempt a quantitative comparison between the microscopic simulations and the macroscopic theory.

But that is not all. As we argued above, the Derjaguin-Anderson theory of diffusio-osmosis ignores the effect of the chemical potential gradient of the solvent. 
However, Eq.~\ref{eq:v_LTE} allows us to go beyond the standard Derjaguin-Anderson theory by taking into account all chemical potential gradients.
Importantly, we can determine the contribution from the different species to the velocity in Eq.~\eqref{eq:v_LTE}. 
In the present case, it is straightforward to estimate the sign of the diffusio-osmotic velocity \textit{a priori}, using thermodynamic arguments~\cite{Anderson1984}. 
However, Eq.~\eqref{eq:v_LTE} can also deal with situations where there is a multi-component solution with competing interactions between the species and the surface.

\begin{figure}[H]
\centering
\includegraphics[width=0.8\linewidth]{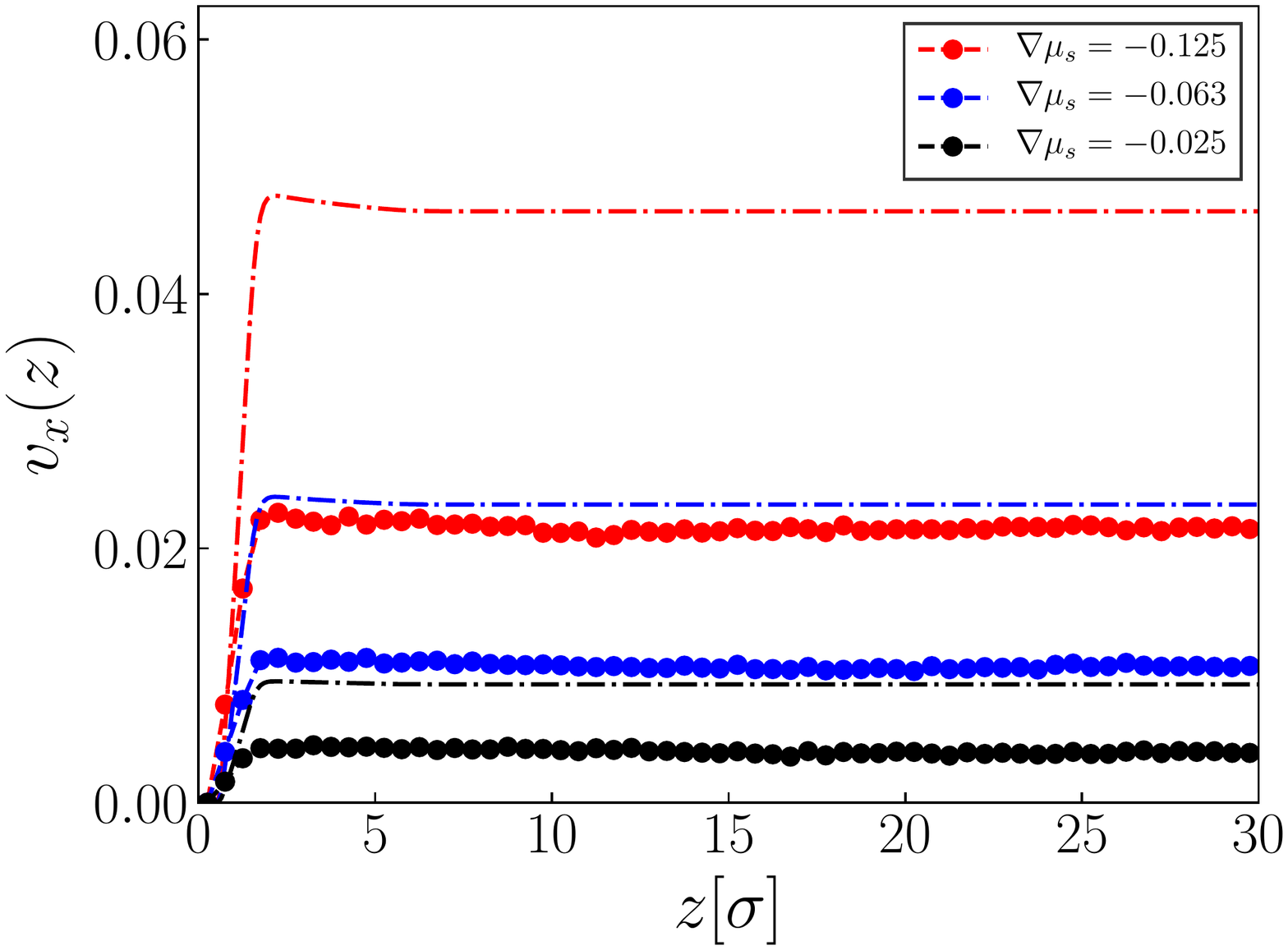}
\caption{Diffusio-osmotic velocity profile for different chemical potential gradients. The simulation results are shown in circles. The theoretical results using Eq.~\eqref{eq:v_LTE} and $z=0$ at a distance $d_f^{\text{min}}$ from the wall, are shown as dash-dotted curves. The velocity sampling bin size is $0.25\sigma^{LJ}_0$. Note that, as explained in the text,  the predictions of the Derjaguin-Anderson theory deviate substantially from the simulations, because the assumptions underlying the theory are not justified on the atomistic scale. }
\label{fig:grad_mu_dependence}
\end{figure}

Fig.~\ref{fig:grad_mu_dependence} shows the velocity profiles for different values of the gradient of the chemical potential $\grad \mu_s$. 
As explained above, the problem with the comparison with the Derjaguin-Anderson theory is twofold: we need to choose the location of the non-slip boundary, and we need to assume constant viscosity.
In Fig.~\ref{fig:grad_mu_dependence} we have computed the theoretical profiles assuming that the non-slip boundary is at $d_f^{\text{min}}$, where the flow velocity vanishes, and we have assumed that the viscosity is everywhere equal to the bulk viscosity.
As the figure clearly shows, with these inputs, there are large discrepancies between theory and simulation.
Of course, a better agreement between theory and the Derjaguin-Anderson theory can be achieved by changing our choice for the local viscosity and the location of the slip plane, but then we would be fitting rather than predicting.
However, even using a local viscosity is not solving the problem, as the viscosity is not a local quantity, as its Fourier transform is wave-vector dependent~\cite{Todd2008a,Oga2019}. 

We also note that the parabolic part of the pressure-driven velocity profile in Fig.~\ref{fig:p_driven_flow} extrapolates to zero at a distance from the wall where the real flow velocity is non-zero. 

Finally, we  note that the Derjaguin-Anderson theory  predicts an small overshoot in the velocity profile, which can be attributed to the  layering  in the fluid close to the wall. 

\section{The effect of finite P\'eclet numbers}\label{sec:Colloid}
In our discussion of diffusio-osmosis, we considered two simulation techniques: on the one hand an approach where explicit (periodic) concentration gradients are imposed (the ``Boundary-driven" NEMD approach) and on the other hand, an approach where we imposed fictitious color fields that reproduce the effect of chemical potential gradients (the ``Field-driven'' NEMD approach). 
Although in principle, both methods are equivalent, we chose to use the FD-NEMD approach, as it is computationally more convenient. However, in some situations, there are large differences in simulations using FD and BD NEMD. 
To be more precise: the two methods are still equivalent in the limit where the gradients vanish, but non-linearities show up much more strongly in the BD-NEMD approach than in the FD method. In this section we discuss an example, namely colloidal diffusiophoresis, where these differences can be shown quite clearly.

\subsection{Boundary-Driven Non-Equilibrium Molecular Dynamics}
For our BD-NEMD, we used a double-control-volume semi-grand canonical algorithm. We use a semi-grand canonical ensemble that allows us to swap particles between the two reservoirs. 
As we consider again solvent and solute particles that are otherwise identical, all swap moves are accepted. 

\begin{figure}[H]
\centering
\includegraphics[scale=0.6]{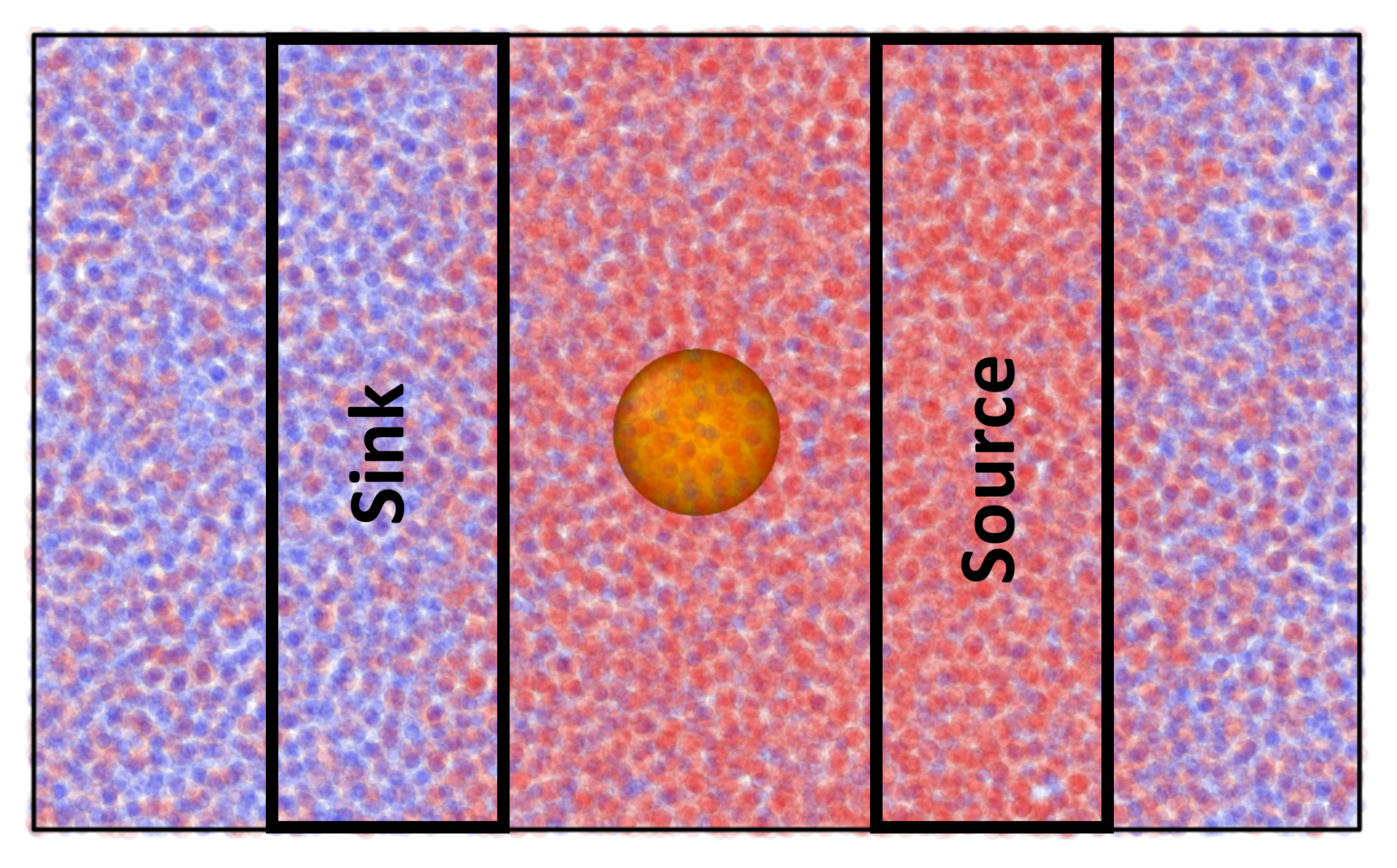}
\caption{Dual control volume simulation box used for the boundary-driven non-equilibrium simulations. In both control volumes, the concentration for each particle species was fixed, with the \textit{sink} and \textit{source} indicating the low and high solute concentration regions respectively. The distance between the reservoirs is $\Delta x^{\text{ss}}=12x_l$ and the length of the control volumes in the $x$ direction is $\Delta_x^{\text{cv}}=3x_l$, where $x_l=5^{1/3}\sigma_0$.}  
\label{fig:bd_system}
\end{figure}

 The size of the simulation box was (51.30 x 20.52 x 30.78) (in units of ${\sigma}_0$). 
 A colloid was fixed in the centre of the simulation box (see Fig.~\ref{fig:bd_system}) by placing a large Lennard-Jones particle with $\sigma_{cs} = \sigma_{cf}$ = 3.23 $\sigma_0 $, where the subscript $c$ denotes the colloid. The concentration gradient was created by using two reservoirs of particles. The \textit{source} region at $c_s^B=0.6 \sigma^{-3}_0$ and the  \textit{sink} at $c_s^B=0.15 \sigma^{-3}_0$. The difference in concentration between the reservoirs is equivalent to $\nabla \mu_s \sim 0.06$. The imposed concentration gradient is linear when the interaction of the colloid with solvent and solute is the same:  $\varepsilon_{cs} = 1.0$ (see Fig~\ref{fig:conc_chunks}). 
 Note that, rather than probing the steady-state notion of a colloid in a stationary fluid, we compute the (equivalent) steady state flow of the fluid past a fixed colloid.

As before, the MD simulations were carried out using LAMMPS, and with the same model for solvents ($f$) and solutes ($s$).
Moreover, $\sigma_{cs} = \sigma_{cf}$ = 3.23 and $\varepsilon_{cf}=1$.
The only difference is that we vary the interaction strength between colloid and solute $\varepsilon_{cs}$ to reproduce solute depletion ($\varepsilon_{cs}=0.5$) and attraction ($\varepsilon_{cs}=2.5$).

We initialized the system with a solute/solvent ratio $c_s^B/c_f^B=1$ and an average solution density in the box of $\bar{c}=0.75 \sigma_0^{-3}$. 
We swapped particle identity in the reservoirs every 20 time steps, with a time step of $\Delta t = 0.05\tau$. 
We let the system equilibrate for $10^7$ steps. 
By doing this, we achieve both  equilibration of solutes around the colloid and the desired concentration gradient between the control volumes. 
The equations of motion were integrated using a velocity-Verlet algorithm, and we kept the temperature of the system at $k_BT/\epsilon_0=1.0$  using a Nos\'{e}-Hoover thermostat \cite{Hoover1985}. 
After  equilibration, we ran $10^7$ production steps to sample the flow profile around the colloid and the concentration distribution for each species.

\subsection{BD-NEMD Results}
\begin{figure}[H]
\centering
\includegraphics[scale=0.85]{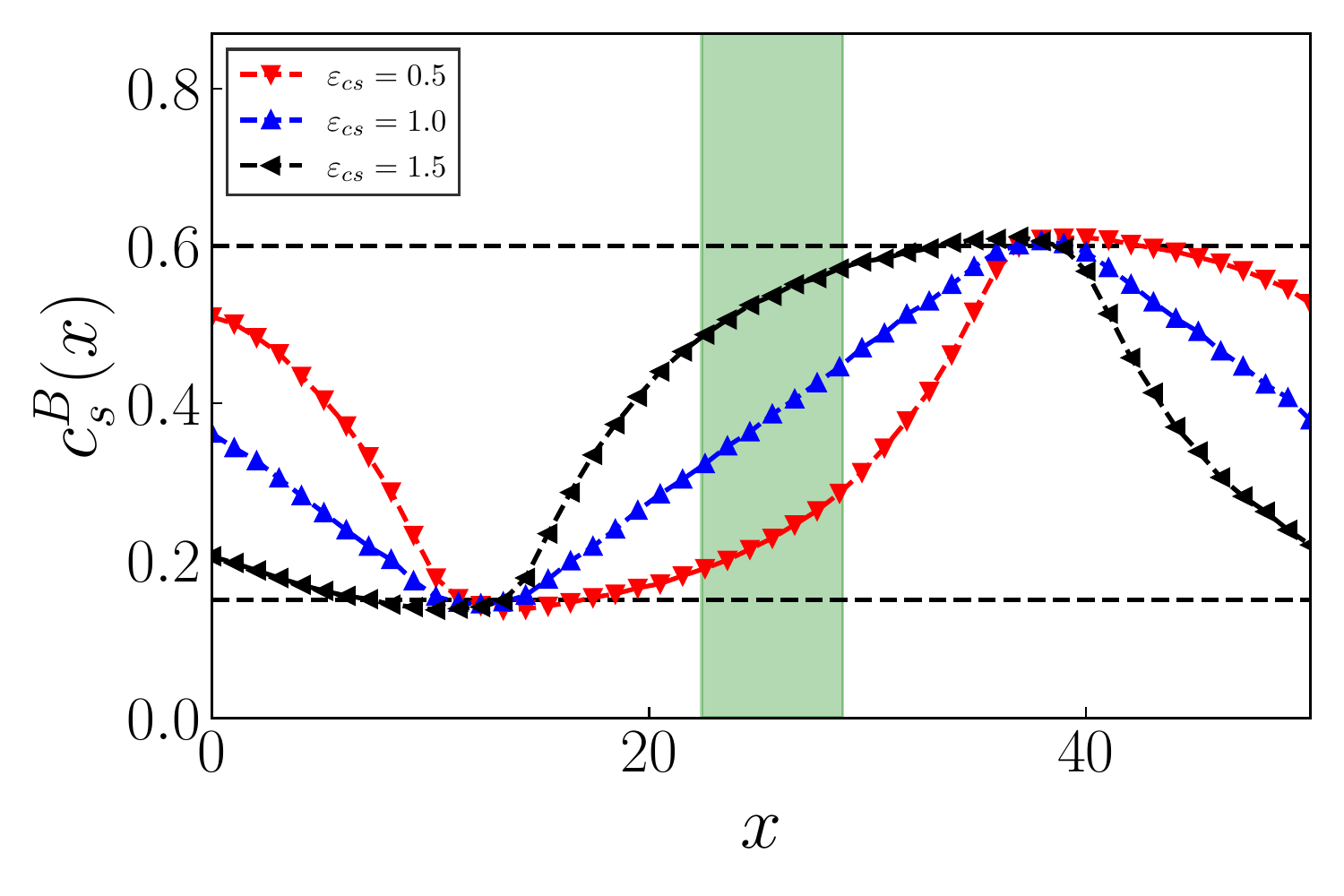}
\caption{Solute concentration profile for a constant concentration gradient. We show the results for different phoretic flow velocities, corresponding to several values of  $\varepsilon_{cs}$. 
We measure the concentration profiles at a lateral distance of at least 10$\sigma$ from the colloid, where the colloid does not directly perturb the concentration profile. 
The shaded region represents the $x$ position of the colloid, which we show to emphasize the asymmetry in the concentration distribution created by the advection.}  
\label{fig:conc_chunks}
\end{figure}

In Fig.~\ref{fig:conc_chunks} we show the solute concentration profiles for different colloid-solute interactions $\varepsilon_{cs}$. As soon as phoresis starts, i.e. for $\epsilon^{LJ}_{cs} \neq 0$, the concentration gradient becomes non-linear due to advection. 
This is a consequence of the finite P\'eclet number, which in this case is of the order of $Pe\approx vL/D$, where $v$ is the average flow velocity, $L$ the distance between the reservoirs, and $D$ the diffusion coefficient of both solute and solvent. 
As a result, the local concentration gradient at the location of the colloid decreases (see also \cite{Wei2020}). 

In fact, a simple argument shows that  the concentration profile should become approximately  exponential. 
To this end, we consider the fluxes of solvent and solute in a steady velocity field $\vb*v$. 
We ignore the fact that the colloid presents an obstacle. 
We can then write the flux of species $i$ as a sum of a diffusive  and a convective contribution,
\eq{\vb*{J}_i=-D_i\grad c_i^B+\vb*v c_i^B\;.} 
In steady state, the concentration profile must be of the form
\eq{c_i(x)=\alpha\;e^{(v/D_i) x}+\beta\;,}
where $D_s/v=1/k$ defines the  characteristic length scale of the concentration profile.  The coefficients $\alpha$ and $\beta$ are determined by the boundary conditions at the source and sink regions of the system (see Fig.~\ref{fig:bd_system}).

\begin{figure}[H]
\centering
\includegraphics[width=0.6\linewidth]{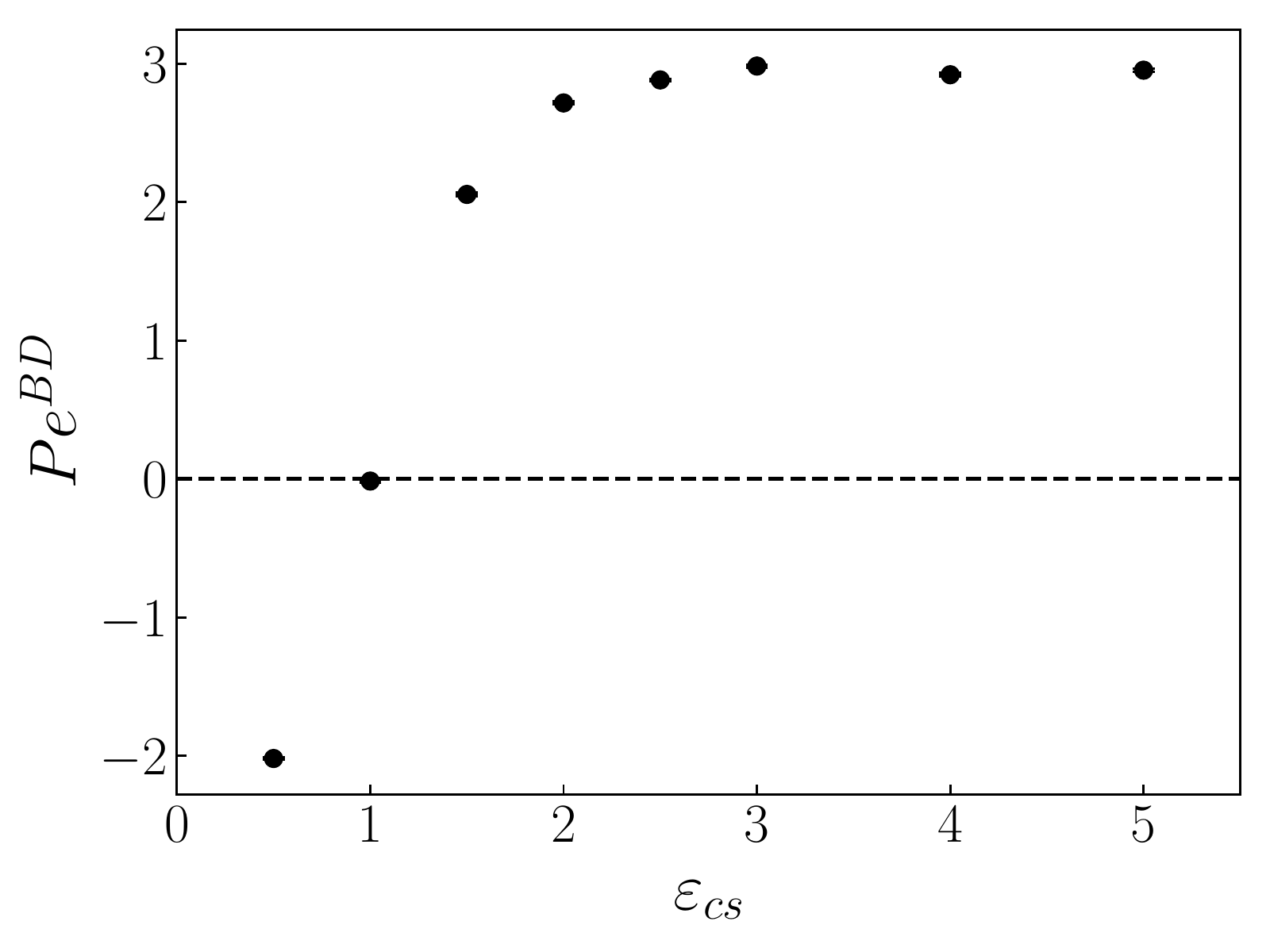}
\caption{P\'eclet number $Pe^{BD}$ for the diffusiophoretic flow with several colloid-solute interaction strengths $\varepsilon_{cs}$. Note that, even for the smallest non-zero phoretic flow velocities $Pe^{BD}$ is larger than one.}
\label{fig:Peclet}
\end{figure}

If we restrict the analysis to the solutes and set $c_s(0)=c_s^\text{sink}$ and $c_s(\Delta x^{\text{ss}})=c_s^\text{source}$, with $\Delta x^{\text{ss}}$ being the distance between the control volumes, we have:

\eq{c_s^B(x) = c_s^\text{sink}+\frac{e^{kx}-1}{e^{k\Delta x^{\text{ss}}}-1}\Delta c_s^B\;,}
with $\Delta c_s^B = c_s^\text{source}- c_s^\text{sink}$. We can define the P\'eclet number for the BD-NEMD simulations as $Pe^{BD} =k\Delta x^{\text{ss}}$ =  $v\Delta x^{\text{ss}}/D_s$. 
In Fig.~\ref{fig:Peclet} we show $Pe^{BD}$ for the different interactions $\varepsilon_{cs}$. $D_s=0.13$ was computed from the mean-square displacement of the solutes in the bulk region.  
Even for the smallest non-zero phoretic flow $Pe^{BD}$ is  non-negligible, and the BD-NEMD simulations cannot be used to estimate diffusiophoresis.

For freely moving colloids,  the effect of the finite P\'eclet number on the speeds of diffusiophoresis is well known~\cite{Khair2013,Anderson1984}.
However, the large effect of a finite P\'eclet number in simulations with the boundary-driven flow seems less known.
In fact, Sharifi {\em et al.}~\cite{Sharifi-Mood2013} reported BD-NEMD simulations of diffusiophoresis, but in order to suppress the P\'eclet effect, they were forced to make the concentration profile piece-wise linear, which introduces unphysical sources and sinks in the diffusive fluxes throughout the simulation box

\subsection{Field-Driven Non-Equilibrium Molecular Dynamics}
\label{section:colloid_FD}

To carry out FD-NEMD simulations of the same model system, we used a simulation box (20.52 x 20.52 x 30.78) (in units of ${\sigma}_0$). The system was initialized using  the same procedure as in the BD-NEMD simulations. 
To equilibrate, in this case, we imposed semi-grand canonical swap moves between $s$ and $f$ throughout the simulation  box. 
We attempted to swap $10^4$ particle identities every $10$ steps for the first $10^5$ steps, thereby generating an equilibrium distribution of solutes around the colloid, and an equimolar solution in the bulk. 
The equilibration step is crucial as our aim to carry out simulations under conditions where the composition of the bulk fluid is kept fixed, even as we varied the colloid-solute interaction $\varepsilon_{cs}$~\cite{Ramirez-Hinestrosa2019}.

\subsection{FD-NEMD Results}
As discussed previously, we represent the chemical potential gradients by equivalent external forces that are compatible with the periodic boundary conditions. 
As before, the forces are chosen so that there is no net force on the system as a whole.
Hence, there is only one independent force to be chosen.
 In the present case, we chose to fix the force on the solutes ${F}_{s}^{\mu}$. 
 To facilitate comparison with the BD-NEMD simulations, we fixed this force such that it corresponds to a linear concentration gradient in the BD-NEMD case.  
 This choice resulted in, ${F}_{s}^{\mu}=0.06 \epsilon_0/\sigma_0$. Following the discussion in
 Sec.~\ref{sec:ch_do_simulations} and bearing in mind the complex geometry of the present case, we applied color forces on the solvent and solute, rather than average forces on the fluid as in diffusio-osmosis (see Eq.\eqref{eq:AverageForce_mu}). 

Having specified the force on the solutes, the force on the solvent particles $\mathbf{F}_{f}^{\mu}$ follows from mechanical equilibrium in the bulk:

\eq{\label{eq:bulk_equilibrium_col}\mathbf{F}_{s}^{\mu} N_s^B +\mathbf{F}_{f}^{\mu}  N_f^B  = 0, }
$N_s^B$, $N_f^B$ denote the number of solutes and solvent in the bulk region. Once the forces in the bulk are specified, we obtain the phoretic force on the colloid $\mathbf{F}_{c}^{\mu}$ by imposing mechanical equilibrium in the whole system,

\eq{\mathbf{F}_{c}^{\mu}=-(\mathbf{F}_{s}^{\mu}N_s+\mathbf{F}_{f}^{\mu}N_f), \label{eq:colloid_force}}
$N_s$, $N_f$ refer to the number of solutes and solvents in the whole system. This equation expresses the fact that there can be no net external force on the fluid: if there were, the system would accelerate without bound, as there are no walls or other momentum sinks in the system. Eq.~\eqref{eq:colloid_force} establishes a connection between all chemical potential gradients (or the corresponding microscopic forces), which must be balanced throughout the system as the phoretic flow cannot cause bulk flow.

In practice, we exploit Galilean invariance, and keep the position of the colloid fixed.
As discussed before in the context of diffusio-osmosis (Sec.~\ref{subsec:FDNEMD}),  there are inevitably fluctuations in the bulk concentrations due to exchanges between adsorbed and non-adsorbed particles. These variations would lead to unphysical velocity fluctuations in the bulk (unphysical, because in the thermodynamic limit this effect goes away), creating noise in the observed phoretic flow velocity. To suppress this effect, we could either adjust the composition in the bulk domain at every time step or recompute the forces on the solvents $\mathbf{F}_{f}^{\mu}$) such that the external force on the bulk domain is always rigorously equal to zero (this also adjusts the force on the colloid $\mathbf{F}_{c}^{\mu}$). We opted for the latter approach as particle swaps would affect the stability of the MD simulations. 

\begin{figure}[H]
\centering
\includegraphics[width=0.7\linewidth]{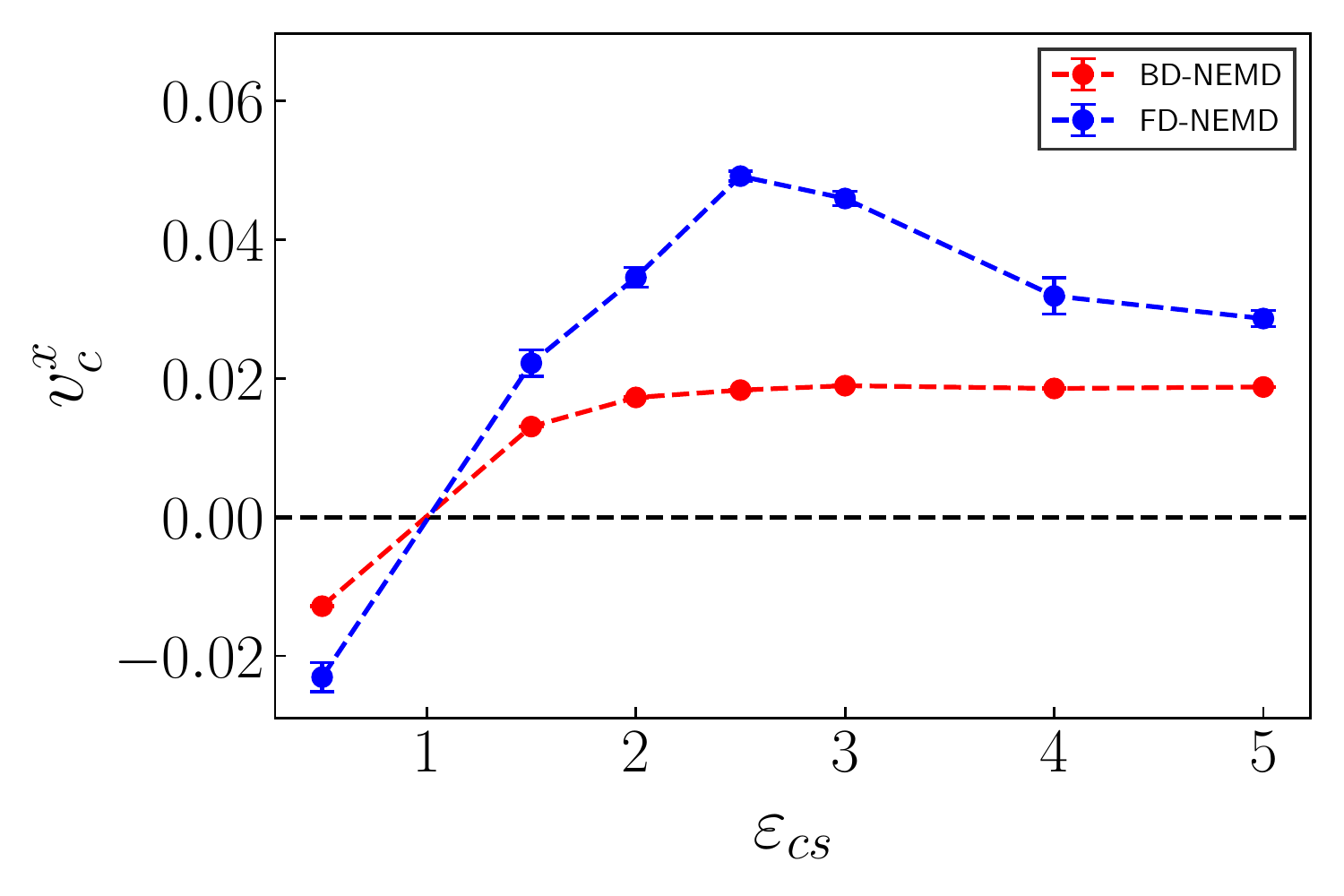}
\caption{Phoretic velocity $v_c^x$ for several colloid-solute interaction strengths $\varepsilon_{cs}$. We compare the results for an explicit concentration gradient (BD-NEMD) and FD-NEMD}  
\label{fig:results_BD_vs_FD}
\end{figure}

In Fig.~\ref{fig:results_BD_vs_FD} we show the results obtained using the BD-NEMD and FD-NEMD. The first point to note is that the BD-NEMD simulations yield a phoretic flow velocity that is systematically lower than the value obtained from the FD-NEMD simulations.
The underlying reason is that whereas the characteristic P\'eclet number in the BD-NEMD case is determined by the system size ($Pe \sim Lv/D$), the P\'eclet number for FD-NEMD is determined by the colloid size ($Pe\sim \sigma_c v/D$, which in our case is about an order of magnitude smaller.

We note that the dependence of the phoretic velocity on the strength of the interaction between colloid and solute is non-monotonic. 
The reason is that initially, $\varepsilon_{cs}$ increases the excess of solutes around the colloid, which, in turn, increases the phoretic velocity as expected in the linear regime. However, for large $\varepsilon_{cs}$, the closest solutes to the colloid are tightly bound and lose their mobility. Hence, they stop contributing to the flow around the colloid (see a discussion in \cite{Ramirez-Hinestrosa2019}).

\section{Conclusions}
In this paper, we have discussed diffusio-osmotic flow from the perspective of (non-equilibrium) thermodynamics and used that framework to define the driving forces in MD simulations. To arrive at a self-contained narrative, the present work contains much that is standard, but not necessarily well-known. We thoroughly presented the connection between non-equilibrium thermodynamics and the microscopic mechanisms driving phoretic flow. Moreover, we generalised Derjaguin-Anderson's theory. Our expression is based on the LTE approach. We include the contributions to diffusio-osmosis from all the species in a multi-component solution. In particular, we take into account the solvent particles. This discrete treatment of the solvent is a crucial difference with respect to previous works based on continuum frameworks.

Our most important conclusion is that it is possible to arrive at a consistent description of diffusio-osmosis and diffusiophoresis in terms of fictitious body forces. 
These forces allow us to carry out Field-Driven non-equilibrium MD simulation (FD-NEMD). 

As much as possible, we have avoided the use of pressure gradients in our description: even though these play a central role in the normal hydrodynamic description, local pressure gradients near walls are ill-defined and their use should be avoided.
In contrast, there seems to be no ambiguity in a description based on chemical potential gradients. 

Of course, simulations of diffusiophoresis can also be carried out using Boundary-driven  non-equilibrium MD (BD-NEMD) and imposing explicit concentration gradients.
However, the BD-NEMD approach runs into practical (although not conceptual) problems in the case of diffusiophoresis, because the system is quickly driven into the regime of non-negligible P\'eclet numbers. 
In fact, this problem is difficult to suppress as the P\'eclet number grows with system size.

The use of fictitious forces is not limited to diffusio-osmosis.
It can also be used in the case of thermo-osmosis. 
However, in that case, the correct description of the field-driven flow is less well-grounded in theory.
For this reason, a comparison between field-driven simulations and boundary-driven simulations of thermo-osmosis with explicit thermal reservoirs~\cite{Joly2021} could help clarify the situation.

\section*{Acknowledgements}
This work was supported by the European Union Grant
No. 674979 [NANOTRANS]. DF acknowledges support from the   Horizon 2020 program through 766972-FET-OPEN-NANOPHLOW. 
SR would like to thank Lyd\'eric Bocquet for his hospitality at the ENS in Paris.  We would like to thank Lyd\'eric Bocquet, Hiroaki Yoshida, Richard Sear, Raman Ganti, Stephen Cox and Patrick Warren for illuminating discussions.
\section*{Author Contribution Statement}
The research was planned by SRH and DF, the simulations and data analysis were carried out by SRH, the manuscript was written by SRH and DF. 
\bibliographystyle{myplain}
\bibliography{library_DF}
\end{document}